\documentclass[aps,preprint,a4paper,showpacs,showkeys,onecolumn,12pt]{revtex4}%
\usepackage{amsfonts}
\usepackage{amsmath}
\usepackage{amssymb}
\usepackage{graphicx}%
\begin{document}
\title{Quantum hypernetted chain approximation for one dimensional fermionic systems}
\author{C\'{e}sar O. Stoico}
\affiliation{Area F\'{\i}sica, Facultad de Ciencias Bioqu\'{\i}micas y Farmac\'{e}uticas,
Universidad Nacional de Rosario, Argentina.}
\author{C. Manuel Carlevaro}
\affiliation{Instituto de F\'{\i}sica de L\'{\i}quidos y Sistemas Biol\'{o}gicos
(IFLYSIB)-CONICET La Plata-UNLP. c.c. 565 (1900) La Plata, \ \ \ \ \ and Universidad Tecnol\'{o}gica Nacional, Facultad Regional Buenos Aires.}
\author{Danilo G. Renzi}
\affiliation{Facultad de Ciencias Veterinarias, Universidad Nacional de Rosario, Casilda, Argentina}
\author{Fernando Vericat}
\altaffiliation{Corresponding author}
\altaffiliation{FAX: +54 221 425 7317}
\altaffiliation{E-mail: vericat@iflysib.unlp.edu.ar}
\affiliation{Instituto de F\'{\i}sica de L\'{\i}quidos y Sistemas Biol\'{o}gicos
(IFLYSIB)-CONICET La Plata-UNLP. c.c. 565 (1900) La Plata, \ \ \ \ \ and Grupo de
Aplicaciones Matem\'{a}ticas y Estad\'{\i}sticas de la Facultad de
Ingenier\'{\i}a (GAMEFI), UNLP, La Plata, Argentina \ \ \ }

\begin{abstract}
In this comprehensible article we develop, following Fantoni and Rosati formalism, a hypernetted chain approximation for one dimensional systems of fermions. Our scheme differs from previous treatments in the form that the whole set of diagrams is grouped: we do it in terms of non-nodal, non-composite and elementary graphs. This choice makes the deduction of equations more transparent.  Equations for the pair distribution functions of one component systems as well as binary mixtures are obtained. We apply they to experimentally realizable quasi-one dimensional systems, the so called quantum wires which we model, within Sommerfeld-Pauli spirit, as a 1D electron gas or as an electron-hole mixture. In order to use our one-dimensional equations we consider pair potentials that depend on the wires width.

\end{abstract}
\keywords{Fermi hypernetted chain approximation, quantum wires, Wigner crystallization,
coupled quantum wires.}
\maketitle

\section{Introduction}

The study of quantum many particle systems amounts to the first half of past
century as a natural development of quantum mechanics. In general it has
evolved along three main lines somehow independent\cite{Mahan1}: one that aims
to describe nuclear matter; other addressed to the so called quantum liquids
($^{3}$He and \ $^{4}$He at low temperatures) and finally a third one devoted
to the electron gas. This last is a simple model to study several properties
of solids (Sommerfeld-Pauli model). More specifically, relative to the third
line, the properties of the ground state of a gas of interacting electrons
have been intensively studied under diverse approximations\cite{Mahan1}. In
general these studies were initially oriented towards the 3D electron gas.
However, the low-dimensional versions of many particles systems in general and
of the electron gas in particular have been receiving an increasing attention.
At the beginning the interest was merely academic\cite{Lieb1},\cite{Mattis1}
but, in last past years, advances in experimental research on organic
metals\cite{Wosnitza1}, carbon nanotubes\cite{Jorio1} and semiconductor
nanostructures\cite{Reed1},\cite{Kelly1} have allowed for true realizations of
some of these systems giving place to more practical motivations to study
them. Today, many body systems in low dimension are very commonly used to
describe real situations in which the movement of the relevant involved
particles is limited in one, two or even three directions (e.g. quantum wells,
quantum wires and quantum dots).

In this work we focus on one dimensional fermionic systems, particularly the
electron gas and mixtures electron-hole as models to study, within
Sommerfeld-Pauli spirit, some of the metallic and semiconductor properties of
quantum wires. Our main tool will be the irreducible diagrammatic formalism
developed by Fantoni and Rosati\cite{Fantoni1} in order to describe 3D many
particle systems. There exist in the literature several
realizations\cite{Fantoni2}-\cite{Lantto2} of Fantoni-Rosati formalism which
differ among them in the way the diagrams are classified and grouped and in
the specific form that the complementary energy\ variational equation
takes\cite{De Dominicis1}, \cite{Zabolitzky1}, \cite{Lantto2}%
,\cite{Krotscheck2}. Here we present our own scheme, adapt it to the one
dimensional case and apply the resulting equations to quantum wires seen as a
1D electron gas or as a 1D mixture of electron and holes in order to study
some of their conductor or semiconductor properties, respectively.

In next Section we briefly show the theory on which we base our calculations,
It is divided into subsections in order to make more clear the presentation.
In the first of these subsections we present the model and the background for
the diagrammatic expansion of the distribution functions within Jastrow
approximation for the many particle wave function. Subsection B is devoted to
to expand the generating function in terms of crude reducible diagrams formed by correlation and exchange lines. 
We also obtain, by functional derivativing the generating function, the diagrammatic expansions for the 
one and two point distribution functions. 
In Subsection C we show how the reducible graphs transforms into irreducible ones and classify them in three classes: 
non-nodal, non-composite and elementary, each one being in turn classified according to the type of lines that converge to their root points.
Two important exact relations among the sums of the diverse class of irreducible diagrams, say
the relation of van Leeuwen \textit{et al.}\cite{van Leeuwen1} and an
Ornstein-Zernike\cite{Ornstein1} like equation are established in subsection D.
Additional relationships derived from the composite structure of the non-nodal graphs are
also presented in this subsection.
All these relations, together with the variational equation for the energy of
Subsection E, define a system of coupled equations for the pair correlation function. 
This system contains the rather difficult to calculate elementary diagrams and, according
to which of them we include in our calculation, we obtain diverse quantum
hypernetted chain approximations named QHNC/\textit{n}, where \textit{n }is
the highest order of the elementary diagrams considered. In this work we will
restrict to the cases $n=0$ (that corresponds to neglecting all the elementary
diagrams) and $n=4$ that takes into account just the smallest order
(elementary diagrams with $4$ vertices). A scheme to approximate the 4th order elemntary graphs is given in subsection F.  
Finally, in Subsection G we outline the 
quantum hypernetted equations for binary mixtures of fermions.

The remainder of the paper (Section III) is used to apply our QHNC equations to
describe quantum wires as one dimensional systems formed by electrons or by
electrons and holes. Strictly speaking, quantum wires are not one-dimensional 
but \textit{quasi }one-dimensional devices, thus in order to
use our description we define effective pair potentials that take into account
the non-zero value of the wire width. Subsection A is reserved to study the
microscopic structure of quantum wires when they are taken as conductors. To
this purpose we model them by a 1D electron gas with effective pair potentials
and calculate pair correlation functions and structure factors. We observe how
these functions change with the density and the wire width. In particular we
note their variation from a liquid like behavior to the one characteristic of a Wigner crystal. 
In order to check the QHNC results we
compare them with variational Monte Carlo calculations. The generalization of
the QHNC equations to mixtures is applied in Section B to quantum wires seen
now as semiconductors, specifically as an electron-hole mixture. We take
advantage of the fact that the pair correlation between an electron and a hole
at contact is a measure of the electron-hole recombination rate to make
contact with photoluminescence experiments. Finally, in Section C, use is made
again of the QHNC equations for mixtures in order to describe the correlations
between the carriers in two parallel coupled quantum wires and show how, in
determined cases, an electron of one of the wires binds to a hole of the other
one so forming a sort of exciton.

\section{Theory}

\subsection{Pair distribution functions}

We consider a system of $N$ fermions (electrons) moving on a segment of the
real axis of length $L$. Actually our interest is in homogeneous, infinite
systems at $T=0$ in the thermodynamic limit ($N\rightarrow\infty$,
$L\rightarrow\infty$ with $N/L\rightarrow\rho$, the constant density). If
$V(x_{i},x_{j})$ is the pair potential between particles, the system
Hamiltonian reads%

\begin{equation}
H=\sum_{i=1}^{N}\dfrac{p_{i}^{2}}{2m}+\sum_{i<j}^{N}V(x_{i},x_{j}),
\label{1}%
\end{equation}
where $p_{i}$ is the momentum of particle $i$ and $m$ the mass of an electron.
Here we restrict ourself to the one component case, the generalization to
mixtures being sketched in Subsection F. The $N$-particles wave function $\psi(x_{1},...,x_{N})$
verifies Schr\"{o}dinger equation $H\psi=E\psi.$ For the ideal gas of fermions
(for which is $V(x_{i},x_{j})\equiv0$) the\ $N$-particles wave function
$\psi_{ideal}(x_{1},...,x_{N})$ is antisymmetric under particles interchange
and can be written%

\begin{equation}
\psi_{ideal}(x_{1},...,x_{N})=\sum_{P}(-1)^{P}P\left\{  \phi_{1}(x_{1}%
).\phi_{2}(x_{2})...\phi_{N}(x_{N})\right\}  =det\left[  \phi_{\alpha_{_{i}}%
}(x_{j})\right]  \label{2}%
\end{equation}
where $P$ is the permutation operator, $\phi_{\alpha_{_{i}}}(x_{j})$ is the
one-particle wave function and $det\left[  \phi_{\alpha_{_{i}}}(x_{j})\right]
$ represents the determinant of the Slater matrix
\begin{equation}
M_{S}=%
\begin{bmatrix}
\phi_{1}(x_{1}) & \phi_{1}(x_{2}) & ... & \phi_{1}(x_{N})\\
\phi_{2}(x_{1}) & \phi_{2}(x_{2}) & ... & \phi_{2}(x_{N})\\
... & ... & ... & ...\\
\ \phi_{N}(x_{1}) & \phi_{n}(x_{2}) & ... & \phi_{N}(x_{N})
\end{bmatrix}
\label{3}%
\end{equation}
Frequently, for the one particle functions are used wave planes of the form%
\begin{equation}
\phi_{\alpha_{i}}(x_{j})=\dfrac{1}{\sqrt{\Omega}}e^{ik_{\alpha_{i}}x_{j}%
}\sigma(j), \label{4}%
\end{equation}
where $\sigma(j)$ denotes the spin contribution and $\Omega$ the volume that
goes to infinity in the thermodynamic limit. The allowed momenta fill a
segment (Fermi "sphere") of "radius" $k_{F}=\pi\rho/\nu$, with $\nu=2$ the
spin degeneration.

For systems in which the particles interact ($V(x_{i},x_{j})\neq0$),
correlations are induced. \ A convenient manner to handle these correlations
is with a trial wave function of the form:
\begin{equation}
\psi(x_{1},...,x_{N})=F(x_{1},...,x_{N})\psi_{ideal}(x_{1},...,x_{N}),
\label{5}%
\end{equation}
The correlation factor $F(x_{1},...,x_{N})$ is assumed to be symmetric under
particles permutations, so the system statistics is determined just by the
anti-symmetric ideal part. An appropriate election for $F$ is the Jastrow
factorization\cite{Jastrow1}
\begin{equation}
F(x_{1},...,x_{N})=\prod_{i<j}f_{2}(x_{i},x_{j}) \label{6}%
\end{equation}
where the two particle correlation factor $f_{2}(x_{i},x_{j})$ goes to zero
when the distance $x_{ij}=|x_{i}-x_{j}|$ is smaller than the range of the
repulsive part of the pair potential and to one for $x_{ij}$ large, denoting
the absence of correlations.

The main objects in our theory are the one and two particles distribution functions:

\begin{equation}
\rho^{(1)}(x_{1})=N\frac{\int dx_{2},...,dx_{N}\psi_{N}^{\dagger}
(x_{1},...,x_{N})\psi_{N}(x_{1},...,x_{N})}{\int dx_{1},...,dx_{N}\psi
_{n}^{\dagger}(x_{1},...,x_{N})\psi_{N}(x_{1},...,x_{N})} \label{7}
\end{equation}
which normalizes such that $\int dx_{1}\rho^{(1)}(x_{1})=N$ \ and%

\begin{equation}
\rho^{(2)}(x_{1},x_{2})=N(N-1)\frac{\int dx_{3},...,dx_{N}\psi_{N}^{\dagger
}(x_{1},...,x_{N})\psi_{N}(x_{1},...,x_{N})}{\int dx_{1},...,dx_{N}\psi
_{N}^{\dagger}(x_{1},...,x_{N})\psi_{N}(x_{1},...,x_{N})}. \label{8}%
\end{equation}
The function $\rho^{(1)}(x_{1})$ is the probability density of finding a particle at $x_{1}$ whereas $\rho^{(2)}(x_{1},x_{2})$ 
is the probability density of finding two particles at $x_{1}$ and $x_{2}$, respectively.

Defining the generating function $G\left\{  U,f\right\}  =\log\left\langle
\psi\right\vert \left.  \psi\right\rangle $ as%

\begin{equation}
G\left\{  U,f\right\} =\log\left\langle \psi_{ideal}\right\vert \prod
_{i<j}\left\vert f(x_{i},x_{j})\right\vert ^{2}%
{\displaystyle\prod\limits_{k=1}^{N}}
\exp U(x_{k})\left\vert \psi_{ideal}\right\rangle , \label{9}%
\end{equation}
where $U\left(  x\right)  $ is an auxiliary function which properly chosen
simplifies calculations, it can be proved that the distribution functions verify%

\begin{equation}
\rho^{(1)}(x_{k})=\frac{\delta G\left\{  U,f\right\}  }{\delta U(x_{k})}
\label{10}%
\end{equation}

\begin{equation}
\rho^{(2)}(x_{i},x_{j})=\frac{\delta^{2}G\left\{  U,f\right\}  }{\delta
U(x_{i})\delta U(x_{j})}+\rho^{(1)}(x_{i})\rho^{(1)}(x_{j})-\delta(x_{i}%
,x_{j})\rho^{(1)}(x_{i}), \label{11}%
\end{equation}
with $\delta(x_{i},x_{j})$ the Dirac delta.

Directly related to the two particle distribution function is the pair
correlation function $g(x_{1},x_{2})$ defined by:%

\begin{equation}
\rho^{(2)}(x_{1},x_{2})=\rho^{(1)}(x_{1})\rho^{(1)}(x_{2})g(x_{1},x_{2}).
\label{12}%
\end{equation}
For homogeneous systems is $\rho^{(1)}(x_{1})\equiv\rho$ and $g(x_{1}%
,x_{2})\equiv g(x_{12})$.

\subsection{Diagrammatic expression for the distribution functions}

In this Subsection we show how the generating function $G\left\{  U,f\right\}
$ can be expressed in terms of graphs whose edges are of two types: ones that
come from the Jastrow factors (correlation or dynamic bonds) and the others coming
from the ideal part of the wave function (exchange or statistical bonds).

Instead of working directly with the Jastrow factors $f(x_{i},x_{j})$ it is
convenient to use the functions $b_{ij}=b_{ij}(x_{i},x_{j})=f^{2}(x_{i}%
,x_{j})-1$ in order to avoid integrals convergence problems. This way , the
product of correlation factors in Eq. (\ref{9}) can be expressed as products
of bond functions $b_{ij}$:%

\begin{align}
\prod_{i<j}\left\vert f(x_{i},x_{j})\right\vert ^{2}  &  =\prod_{i<j}%
(1+b_{ij})=1+\sum_{i<j}b_{ij}+\sum_{i<j}\sum_{k<l}b_{ij}b_{kl}+...\nonumber\\
&  =1+X_{2}+X_{3}+...+X_{N} \label{13}%
\end{align}
where%

\begin{align}
X_{2}  &  =\sum_{i<j}b_{ij}\nonumber\\
X_{3}  &  =\sum_{i<j<k}(b_{ij}b_{jk}+b_{jk}b_{ki}+b_{ki}b_{ij}+b_{ij}%
b_{jk}b_{ki})\label{14}\\
&  \vdots\nonumber
\end{align}

Each term $X_{p}$ can generically be written%

\begin{equation}
X_{p}=\sum_{i<j<...<p}B_{p}(x_{1},x_{2},...,x_{p}) \label{15}%
\end{equation}
with $B_{p}(x_{1},x_{2},...,x_{p})$ being a symmetric function of the
coordinates $x_{1},x_{2},...,x_{p}$. It is convenient to write they in second quantization%

\begin{equation}
X_{p}=\frac{1}{p!}\int dx_{1}...dx_{p}B_{p}(x_{1},x_{2},...,x_{p}%
)\Psi^{\dagger}(x_{1})...\Psi^{\dagger}(x_{p})\Psi(x_{p})...\Psi(x_{1})
\label{16}%
\end{equation}
where the field operators $\Psi^{\dagger}(x)$ and $\Psi(x)$ create and
destroy, respectively, fermions at $x$. Since $X_{p}\left\vert \psi
_{ideal}\right\rangle =0$ for $p>N,$ we see that the finite sum is written as
an infinite one:%

\begin{align}
&  \left\langle \psi_{ideal}\right\vert (1+X_{2}+X_{3}+...+X_{N})%
{\displaystyle\prod\limits_{k=1}^{N}}
\exp U(x_{k})\left\vert \psi_{ideal}\right\rangle =\left\langle \psi
_{ideal}\right\vert
{\displaystyle\prod\limits_{k=1}^{N}}
\exp U(x_{k})\left\vert \psi_{ideal}\right\rangle \nonumber\\
&  +\frac{1}{p!}\sum_{p=2}^{\infty}\int dx_{1}...dx_{p}B_{p}(x_{1}%
,x_{2},...,x_{p})\times\nonumber\\
&  \hspace{3cm}\times\left\langle \psi_{ideal}\right\vert \Psi^{\dagger}%
(x_{1})...\Psi^{\dagger}(x_{p})\Psi(x_{p})...\Psi(x_{1})%
{\displaystyle\prod\limits_{k=1}^{N}}
\exp U(x_{k})\left\vert \psi_{ideal}\right\rangle \label{17}%
\end{align}

The expected values of the field operators can be calculated, using Wick
theorem\cite{Fetter1}, in terms of contractions $\Psi^{\dagger\bullet}%
\Psi^{\bullet}$ which can be written in the form of matrix elements $\rho
_{ij}=\left\langle x_{i}\right\vert \rho(U)\left\vert x_{j}\right\rangle $ of
the density matrix operator $\rho(U)$:%

\begin{equation}
\frac{\left\langle \psi_{ideal}\right\vert \Psi^{\dagger}(x_{j})\Psi(x_{i})%
{\displaystyle\prod\limits_{k=1}^{N}}
\exp U(x_{k})\left\vert \psi_{ideal}\right\rangle }{\left\langle \psi
_{ideal}\right\vert
{\displaystyle\prod\limits_{k=1}^{N}}
\exp U(x_{k})\left\vert \psi_{ideal}\right\rangle }=\left(  \Psi^{\dagger
}(x_{j})\right)  ^{\bullet}\left(  \Psi(x_{i})\right)  ^{\bullet}=\rho_{ij}
\label{18}%
\end{equation}
This way we write the argument of the $\log$ in Eq. (\ref{9}) as:%

\begin{equation}
\left\langle \psi_{ideal}\right\vert \prod_{i<j}\left\vert f_{ij}\right\vert
^{2}%
{\displaystyle\prod\limits_{k=1}^{N}}
\exp U(x_{k})\left\vert \psi_{ideal}\right\rangle =\left\langle \psi
_{ideal}\right\vert
{\displaystyle\prod\limits_{k=1}^{N}}
\exp U(x_{k})\left\vert \psi_{ideal}\right\rangle +\nonumber
\end{equation}%
\begin{equation}
\hspace{2cm}+(\text{Sum of all labeled Jastrow graphs}) \label{19}%
\end{equation}
The parenthesis is the sum of integrals of products of functions
$b_{ij}^{\prime}$s and $\rho_{ij}^{\prime}$s. \ Each of these integrals can be
associated to a labeled Jastrow graph.

A labeled Jastrow graph is a set of $p$ points or vertices labeled by
coordinates $x_{i}$ \ which are linked by correlation bonds $b_{ij}$
(represented by dashed lines) and/or oriented exchange lines $\rho_{ij}$ (here
represented by arrows) in such a way that:

\begin{itemize}
\item each vertex is the extreme of at least one correlation bond $b_{ij}$

\item each pair of vertices is linked at maximum by one bond $b_{ij}$

\item each vertex has one oriented exchange line $\rho_{ij}$ arriving to and
one leaving from it.
\end{itemize}

The contribution of an oriented labeled Jastrow graph is obtained from the
following rules:

\begin{itemize}
\item each bond $b_{ij}$ that links vertices $x_{i}$ and $x_{j}$ contributes
with a factor: $x_{i}---$ $x_{j}=b(x_{i},x_{j})$

\item each oriented exchange line $\rho_{ij}$ leaving vertex $x_{j}$ and
arriving to vertex $x_{i}$ contributes with a factor: $x_{i}\longrightarrow
$--- $x_{j}$ $=\left\langle x_{i}\right\vert \rho(U)\left\vert x_{j}%
\right\rangle =\rho_{ij}$

\item multiply the product of the above factors by $(-1)^{n_{\rho}+n_{L}}/p!$,
where $n_{\rho}$ is the number of the exchange lines, $n_{L}$ the number of
closed loops formed by exchange lines and $p$ is the vertices number.

\item integrate over the coordinates that label the vertices.
\end{itemize}

Each $p$-vertices graph belongs to a family of $p!$ graphs which are obtained,
ones from the others, by permuting the labeled vertices. All graphs belonging
to the same family contribute equal to the total sum. In general -because of
symmetry- among the $p!$ graphs obtained by permutation of the labeled
vertices, there are $S$ identical graphs, so the family is formed by just
$p!/S$ different graphs. The factor $S$ is called the \textit{graphs symmetry
number}. To avoid considering graphs which give the same contribution, it is
customary to just sum distinct non-labeled Jastrow \ graphs multiplied by
$(-1)^{n_{\rho}+n_{L}}p!/S$ . Symbolically::%

\begin{equation}
\left\langle \psi_{ideal}\right\vert \prod_{i<j}\left\vert f_{ij}\right\vert
^{2}%
{\displaystyle\prod\limits_{k=1}^{N}}
\exp U(x_{k})\left\vert \psi_{ideal}\right\rangle =\left\langle \psi
_{ideal}\right\vert
{\displaystyle\prod\limits_{k=1}^{N}}
\exp U(x_{k})\left\vert \psi_{ideal}\right\rangle +\nonumber
\end{equation}%
\begin{equation}
\hspace{2cm}+(\text{Sum of all non-labeled Jastrow graphs}) \label{20}%
\end{equation}

Also, in general, a graph $\Gamma$ can be decomposed into the product of
independent graphs. If the decomposition is in $\nu_{A}$ graphs $\Gamma_{A}$,
$\nu_{B}$ graphs $\Gamma_{B}$,$\cdots$, then the complete graph can be
represented by the product $\Gamma=\frac{\Gamma_{A}^{\nu_{A}}}{\nu_{A}!}%
\frac{\Gamma_{B}^{\nu_{B}}}{\nu_{B}!}\cdots$. The factorials that appear in
this last expression are symmetry factors that correspond to the interchange
of labeled vertices among identical connected parts. We obtain the whole sum
by adding over all the possible values of $\nu_{A}$, $\nu_{B}$, $\cdots$. This
sum is equal to the exponential of the sum of all distinct connected diagrams.
Thus, combining this result with Eq.(\ref{20}) we find that the generating
function $G$, defined by Eq.(\ref{9}) reads%

\begin{equation}
G(U,b)\equiv\log\left\langle \Psi\right.  \left\vert \Psi\right\rangle
=\log\left\langle \psi_{ideal}\right\vert
{\displaystyle\prod\limits_{k=1}^{N}}
\exp U(x_{k})\left\vert \psi_{ideal}\right\rangle +\sum_{K=A,B,\cdots}%
\Gamma_{K}\label{21}%
\end{equation}

According to Eqs.(\ref{10}) and (\ref{11}), in order to calculate the one and
two particles distribution functions from the generating function $G(U,b)$ we
must perform the functional derivatives with respect to $U$. Taking into
account Eq.(\ref{21}), the problem reduces to evaluate derivatives of the kind
$\frac{\delta\Gamma_{K}}{\delta U(x)}$. Since $\Gamma_{K}$ depends on $U$
through a exchange line that contributes with a factor $\left\langle
x_{1}\right\vert \rho(U)\left\vert x_{2}\right\rangle $, we have:%

\begin{align}
&  \frac{\delta\Gamma_{K}}{\delta U(x)}=\int dx_{1}dx_{2}\frac{\delta
\Gamma_{K}}{\delta\left\langle x_{1}\right\vert \rho(U)\left\vert
x_{2}\right\rangle }\frac{\delta\left\langle x_{1}\right\vert \rho
(U)\left\vert x_{2}\right\rangle }{\delta U(x)}\nonumber\\
&  =\int dx_{1}dx_{2}\frac{\delta\Gamma_{K}}{\delta\left\langle x_{1}%
\right\vert \rho(U)\left\vert x_{2}\right\rangle }\left[  \delta
(x_{1},x)\left\langle x\right\vert \rho(U)\left\vert x_{2}\right\rangle
-\left\langle x_{1}\right\vert \rho(U)\left\vert x\right\rangle \left\langle
x\right\vert \rho(U)\left\vert x_{2}\right\rangle \right]  \label{22}%
\end{align}
The right hand side can be more pictorially written in the form %

\begin{equation}
\frac{\delta\Gamma_{K}}{\delta U(x)}=\int dx_{2}
\begin{smallmatrix}
\\
 \includegraphics[width=0.15\textwidth,bb=1 17 105 77]{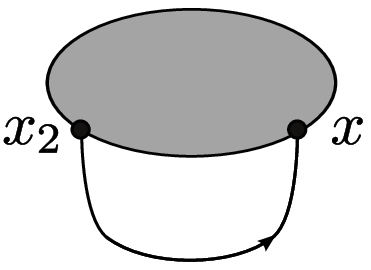}
\\
\\
\end{smallmatrix}
+\int dx_{1}dx_{2}%
\begin{smallmatrix}
\\
\includegraphics[width=0.15\textwidth]{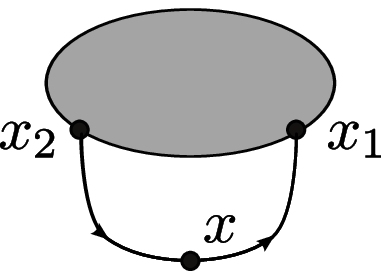}\\
\end{smallmatrix}
\label{23}%
\end{equation}

Thus, equations (\ref{10}) and (\ref{11}) say that the one and two particle distribution 
functions $\rho^{1}(x)$ and $\rho^{2}(x,y)$ are equal to the sum of all connected diagrams with one and two external points $x$ and $(x,y)$, respectively. 
The first diagrams for $\rho^{1}(x)$ and $\rho^{2}(x,y)$ look as:
\begin{eqnarray}
\rho^{(1)}(x) &=& 
\begin{smallmatrix}
 \\
 \includegraphics[width=0.70\textwidth]{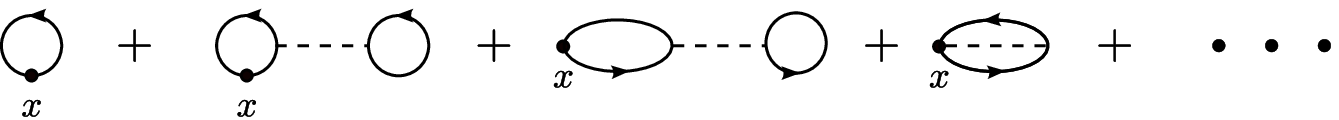}\\
 \\
\end{smallmatrix}     \label{f29}
\\
\rho^{(2)}(x,y) &=& \rho^{(1)}(x)\rho^{(1)}(y)+      \nonumber  \\       
&&\begin{smallmatrix}
                \\
                \includegraphics[width=0.8\textwidth]{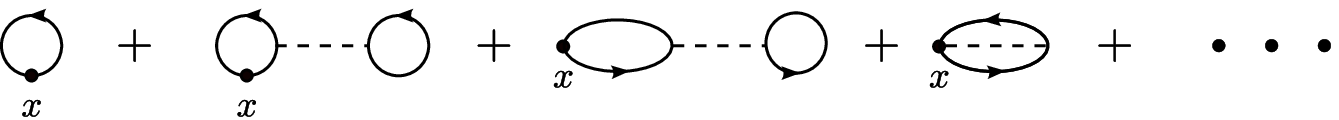}\\
                \\
               \end{smallmatrix}        \nonumber 
 \\ \label{f29b}
\end{eqnarray}

\subsection{Reduction to and classification of irreducible graphs}
Among the graphs that appear in Eqs. (\ref{f29}) and (\ref{f29b}) there are some that have articulation points. Removal of an articulation point from a connected graph causes the diagram to separate into two or more components, of which at least one contains no root points. A graph that is free of articulation points is said to be irreducible. The following generic graph 
\begin{equation}
 \centering
 \includegraphics[width=0.45\textwidth]{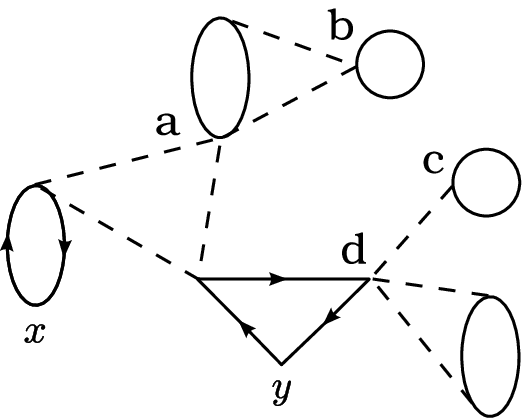} \nonumber
 \label{fig2.1} 
\end{equation}
has four articulation points labeled a, b, c and d. By successively eliminating the articulation points of a given graph, it is always possible to identify a part that contains all the root points. We call this part the irreducible part. Any diagram with articulation points can be reduced to an irreducible one if, in the irreducible part, we associate certain functions to those vertices that were articulation points. These functions ``replace'' those parts of the graph which were eliminated. For the diagram shown above the irreducible part is:
\begin{equation}
 \centering
 \includegraphics[width=0.4\textwidth]{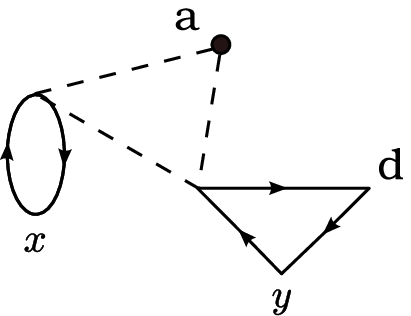} \nonumber
 \label{fig2.2}
\end{equation} 
The points in this graph which were articulation points are or intersections of correlation bonds (like a in the figure) or 
intersections of exchange lines (like d in the figure). We will denote with a black point those which are the intersection of correlation bonds. Given the irreducible part of a graph with a black point, we can assume it can be obtained by reducing any of the graphs $\Gamma$ which can be formed by superimposing to the black point the root point of any graph $\Gamma^{'}$ that has just one root point. The sum of all these graphs $\Gamma^{'}$ is, according to Eq. \ref{f29}, $\rho^{(1)}(x)$. Thus we replace the sum of all the diagrams $\Gamma$ by the irreducible part that is common to all of them and whose black point has the density $\rho^{(1)}(x)$ associated.
Analogously, given an irreducible graph which has a field point that were an articulation one and which is the intersection of two exchange lines (arrows), we can assume it can be obtained by reducing any of the graphs $\Delta$ which can be formed by superimposing to the field point the root point of any graph $\Delta^{'}$ that has just one root point but with the condition that this point be one of the extremes of a correlation bond. Let $A(x)$ be the sum of all the graphs $\Delta^{'}$, it can be seen (c.f. \cite{Hansen1} page 85) that
\begin{equation}
A(x)=\exp U_{int}(x)-1  \label{f31}
\end{equation}
where $U_{int}(x)$ is the sum of all those graphs that belong to the set whose sum is $A(x)$ but which are not products of graphs in the set.
It can be demonstrated that the sum of all reducible graphs whose irreducible parts differ in the number of interchanges lines of some subdiagram \textit{S} (as is exemplified in the following figure)
\begin{equation}
 \centering
 \includegraphics[width=1.0\textwidth]{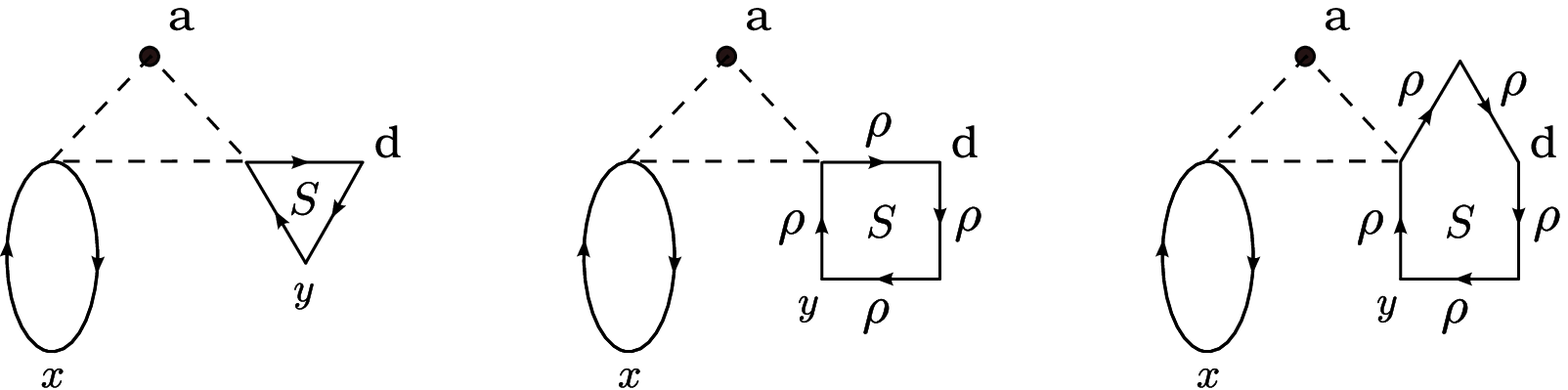} \nonumber
 \label{fig2.3}
\end{equation}
equals the value of just one irreducible graph (see next figure) where now \textit{S} corresponds to the limiting case of infinite interchange lines that can be though as just a line with associated function $\widetilde{\rho }_{ij}$. 
\begin{equation}
 \centering
\includegraphics[width=0.3\textwidth]{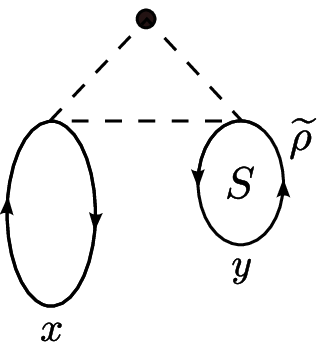} \nonumber
\label{fig2.4}
\end{equation}

Here $\widetilde{\rho }_{ij}$ denotes the elements of matrix $\widetilde{\rho }$ which replace the elements $\rho_{ij}$ associated to each of the previous $n$ interchange lines ($n =3,4,...$). We have
\begin{eqnarray}
\widetilde{\rho }&=&\rho +\rho A-\rho A\rho -\rho A\rho A\ldots \nonumber \\
&=&\rho +\rho A-\rho A\widetilde{\rho }  \nonumber  \\
&=&\rho (U+U_{int})
\label{f34}
\end{eqnarray}
From the last equality we see that, by choosing $U(x)=-U_{int}(x)$, the contribution $\widetilde{\rho}$ of the interchange lines in the irreducible diagrams equals $\rho(0)$. Thus, there are no more unknown functions and the distribution functions $\rho^{(1)}(x)$ and $\rho^{(2)}(x,y)$ are formally expressed as the sum of all the connected irreducible diagrams without articulation point linked to root points.
However, although we know how to evaluate any single diagram, since we are treating with infinite sums, to obtain $\rho^{(1)}(x)$ and $\rho^{(2)}(x,y)$ by simply adding one by one the diagrams so calculated is not practical at all. It is convenient to find a way to globally sum all the involved diagrams into a closed form. To this end, here we classify the irreducible graphs into three classes. In turn the diagrams belonging to any of these classes, except those that we call linear graphs, can be classified according to the kind of lines that arrive or leave their root points. It should be mentioned that the linear graphs are connected irreducible diagrams that have two root points with just one exchange line leaving from one of them and also just one arriving to the other one. All the linear graphs belong to the same group.

In what follows, we focus on the two point distribution function. Thus all the graphs we will be concerned with are Jastrow irreducible graphs with two root points. According to Van Leeuwen, Groeneveld y DeBoer\cite{van Leeuwen1} we classify these diagrams into nodal, composite and elementary graphs with respect to the root points.
We say that a diagram is \textit{nodal} (\textit{non-nodal}) with respect to its root points
if it can (can not) be separated into parts with the two root points appearing in different components.

We call a diagram \textit{composite} (\textit{non composite}) with respect to a couple of points $1$ and $2$ if it is (it is not) composed of two or more parts such that the only points they share are $1$ and $2$.
 
It is straightforward to see that any \textit{nodal} diagram is \textit{non composite} and that every \textit{composite} diagram is \textit{non nodal}.

A graph which is both \textit{non nodal} and \textit{non composite} with respect to the points $1$ and $2$  is said to be \textit{elementary} with respect to those points. Clearly an \textit{elementary} diagram must have at least four (root plus field) points. It is convenient here to define the \textit{order} of a graph as the total number of points it has. Thus, the smallest order of the elemental graphs is $4$.

According to the kind of lines that arrive to their root point, we will group diagrams into three groups that will be indexed
$(bb)$, $(be)=(eb)$ and $(ee)$ ($b$ for correlation bonds and $e$ for exchange lines). A root point is called $b$ if only correlation lines converge to it and it is called $e$ if there exist an exchange line arriving and leaving the point (to which can additionally converge -or not- correlation lines).
A graph is of type $(i,j)$ with $i,j=b,e$ if its root points are of kind $i$ and $j$, respectively.

We denote with $A_{ij}$ the sum of all the diagrams of class $A$ and type $(i,j)$. We will use $A=N$ for the non-nodal graphs; $A=C$ for the non-composite and $A=E$ for the elementary ones. For example, $N_{be}$ represents the sum of all  non-nodal diagrams of type $(be)$. Thus is convenient to use the matricial notation ($A=N,C,E$):

\begin{equation}
A(x,y)=\left( 
\begin{array}{cc}
A_{bb}(x,y) & A_{be}(x,y) \\ 
A_{eb}(x,y) & A_{ee}(x,y)
\end{array}
\right)  \label{c2b}
\end{equation}

\subsection{Hypernetted chain equations}
The sum of all connected and irreducible diagrams with root points $x$ y $y$ of kind $(ij)$  ($i,j=b,e$) will be denoted $L_{ij}(x,y)$.
The following result of Van Leeuwen, Groeneveld y DeBoer\cite{van Leeuwen1} relates in matricial form the non-nodal, non-composite and elementary graphs: 
\begin{equation}
N(x,y)=E(x,y)+b(x,y)I_{bb}+L(x,y)-C(x,y)  \label{teorema}
\end{equation}
where $I_{bb}$ is the $2\times2$ matrix which has the element $(bb)$ one and all the other zero.

All the graphs we call linear belong to just one kind. We denote with $l(x,y)$ the sum of all the linear  diagrams and with $n(x,y)$, $c(x,y)$ and $e(x,y)$ the sum of all the linear diagrams that belong to the non-nodal, non-composite and elementary classes, respectively. They also verify the Van Leeuwen, Groeneveld y DeBoer relation:

\begin{equation}
n(x,y)=e(x,y)+l(x,y)-c(x,y)  \label{teorema2}
\end{equation}

We show now another relationship among the three classes of irreducible diagrams which, together with Eq.
(\ref{teorema}) are the basis of a number of liquid state theories:

\begin{equation}
L(x,y)=N(x,y)+\int dzN(x,z)J(z)L(z,y)  \label{c5}
\end{equation}
where
\begin{equation}
J(x)=\left( 
\begin{array}{cc}
\rho^{(1)}(x) & 1 \\ 
0 & 0
\end{array}
\right)  \label{c6}
\end{equation}

In liquid theory Eq. (\ref{c5}) is usually called the Ornstein-Zernike relation (O-Z)\cite{Hansen1}.
The form of matrix $J(x)$ is due to the fact that two non-nodal diagrams of the type  - - - - - - -   can only be linked in series by a field point which has a factor $\rho^{(1)}(x)$ associated.
Analogously the sum of linear diagrams $l(x,y)$ can be expressed in terms of $n(x,y)$: 
\begin{equation}
l(x,y)=\rho^{(2)}(x,y) +\left[ 1-\rho^{(2)}(x,y )\right]n(x,y)\left[ 1- l(x,y)\right]   \label{c8}
\end{equation}.

On the other hand, for an homogeneous system so that $\rho^{(1)}(x)=\rho=N/L$, Eq. (\ref{f29b}) is written in terms of $L_{ij}$ ($i,j=b,e$) in the form:
\begin{equation}
\rho^{(2)}(x_{12})-\rho^{2} = \rho^{2}\left[ g(x_{12})-1\right] = \rho^{2}L_{bb}(x_{12})+\rho L_{be}(x_{12})+\rho L_{eb}(x_{12})+L_{ee}(x_{12})  \label{h2}
\end{equation}

If the matrix $E$ of elementary diagrams is known then, replacing into Eq.(\ref{c5}) the non-nodal matrix $N$ by the expression given in Eq.(\ref{teorema}), we have a matricial integral equation which would allow to calculate, except for the fact that matrix of non-composite graphs $C$ remains unknown, the matrix $L$ and so, according to Eq.(\ref{h2}), $g(x_{12})$. In order to determine the non-composite diagrams we consider the composite structure of the non-nodal diagrams. This way we obtain additional relations among non-nodal, non-composite and elementary graphs. These relations are not of matricial kind but they depend on the particular type of the non-nodal graphs:  

\begin{align}
N_{bb}(x_{12})=-1-C_{bb}(x_{12})+E_{bb}(x_{12})&+b(x_{12})+f^{2}(x_{12})\exp \left[C_{bb}(x_{12})-b(x_{12})\right] \label{p1}  \\
N_{be}(x_{12})=E(x_{12})+L_{bb}(x_{12})C_{be}(x_{12}&)  \\
N_{ee}(x_{12})=E_{ee}(x_{12})-l(x,y)c(x_{12})+&C_{be}(x_{12})L_{eb}(x_{12})+C_{ee}(x_{12})L_{bb}(x_{12}) \\ 
n(x_{12})=e(x_{12})+l(x_{12})c(x_{12})\hspace{0.75cm}&  \label{p4}
\end{align}

Using Eqs. (\ref{teorema2}), (\ref{c8}) and (\ref{p4}) the sums of linear graphs $l(x_{12})$, $n(x_{12})$ and $c(x_{12})$ are eliminated in favor of the sum of the elementary linear diagrams $e(x_{12})$ which we are assuming as known.

The relations of Van Leeuwen, Groeneveld y DeBoer (\ref{teorema}), (\ref{teorema2}); the O-Z relations (\ref{c5}), (\ref{c8}) and the non-nodal structure relations (\ref{p1})-(\ref{p4}) constitute our hypernetted chain equations.

It remains to know the Jastrow factor $f^{2}(x_{12})$. It can be calculated from the energy variation. A convenient way to proceed is to use the hypernetted chain equations in order to eliminate $f^{2}(x_{12})$ in favor of $g(x_{12})$. We define
\begin{eqnarray}
D &=&L-N  \label{h3}  \nonumber \\
d &=&l-n-\rho, 
\end{eqnarray}
so that $D$ and $d$ mean the sum of all the graphs with two root points that have at least one nodal point and use
\begin{equation}
g_{B}(x)=1+L_{bb}(x)=f^{2}(x)\exp \left[ E_{bb}(x)+D_{bb}(x)\right]
\label{h4}
\end{equation}
to algebraically eliminate the functions $f^{2}$, $N_{bb}$ y $D_{bb}$:

\begin{eqnarray}
f^{2}(x) &=&g_{B}(x)\exp \left[ -E_{bb}(x)-D_{bb}(x)\right]  \label{h5}  \nonumber \\
\widetilde{N}_{bb}(q) &=&-\frac{1}{\widetilde{S}(q)}+\frac{\left[ 1-\widetilde{N}_{be}(q)\right] ^{2}}{%
1+\widetilde{N}_{ee}(q)}  \nonumber \\
\widetilde{D}_{bb}(q) &=&\frac{\left( \widetilde{S}(q)-1\right) ^{2}}{\widetilde{S}(q)}+\frac{
2\widetilde{N}_{ee}(q)+2\widetilde{N}_{be}(q)-\widetilde{N}_{be}^{2}(q)}{1+\widetilde{N}_{ee}(q)}-  \nonumber \\
&&\widetilde{S}(q)\frac{\left[ \widetilde{N}_{ee}(q)+\widetilde{N}_{be}(q)\right] \left[
2-\widetilde{N}_{be}(q)+\widetilde{N}_{ee}(q)\right] }{\left[ 1+\widetilde{N}_{ee}(q)\right] ^{2}}.
\end{eqnarray}
where $\widetilde{S}(q)$ and $\widetilde{N}_{ij}(q)$ $(ij = bb, be, ee)$ denote the Fourier transforms of $h(x)=g(x)-1$ and $N_{ij}(x)$, respectively.

Therefore our QHNC equations are written in the form:
\begin{eqnarray}
\widetilde{D}_{be}(q) &=&-\widetilde{N}_{be}(q)+\widetilde{S}(q)\frac{\left[ 1-\widetilde{N}_{be}(q)\right] \left[
\widetilde{N}_{ee}(q)+\widetilde{N}_{be}(q)\right] }{\left[ 1+\widetilde{N}_{ee}(q)\right] ^{2}}-\frac{\widetilde{N}_{ee}(q)
}{1+\widetilde{N}_{ee}(q)}  \label{h6} \nonumber \\
\widetilde{D}_{ee}(q) &=&-\frac{\widetilde{N}_{ee}^{2}(q)}{1+\widetilde{N}_{ee}(q)}+\widetilde{S}(q)\frac{\left[
\widetilde{N}_{ee}(q)+\widetilde{N}_{be}(q)\right] ^{2}}{\left[ 1+\widetilde{N}_{ee}(q)\right] ^{2}}  \nonumber
\\
\widetilde{d}(q) &=&\left\{ 
\begin{array}{cc}
-\widetilde{n}(q) & \text{para }\left| q\right| <k_{F} \\ 
-\frac{\widetilde{n}^{2}(q)}{1+\widetilde{n}(q)} & \text{para }\left| q\right| >k_{F}
\end{array}
\right.  
\label{h7a}  
\end{eqnarray}
\begin{eqnarray}
N_{be}(x) &=& g_{B}(x)\left[ E_{be}(x)+D_{be}(x)\right] -D_{be}(x)  \nonumber  \\
N_{ee}(x) &=& g(x)-D_{ee}(x)  \nonumber \\
n(x) &=& g_{B}(x)\left[ e(x)+d(x)+\frac{\sigma (x)}{\nu }\right] -d(x)-\frac{\sigma (x)}{\nu }  \nonumber \\
g(x) &=& g_{B}(x)\left\{E_{ee}(x)+D_{ee}(x)+\left[E_{be}(x)+D_{be}(x)\right] ^{2}\right. \nonumber 
\end{eqnarray}
\begin{equation}
\left. -\nu \left[ e(x)+d(x)+\sigma (x)/\nu\right] ^{2}\right\} \\
\label{h7}  
\end{equation}
where $\nu $ denotes the spin degeneration of particles and $\sigma(x)$ is Slater function.

\subsection{Variational equation for energy}

Eqs. (\ref{h7a}) and (\ref{h7}) is a system of seven equations for the 8 unknowns $N_{be}(x)$, $N_{ee}(x)$, $n(x)$, $D_{be}(x)$, $D_{ee}(x)$, $d(x)$, $g_{B}(x)$ and $g(x)$ so that an additional equation is needed. As such we consider the one we obtain from the variation of energy with respect to $g(x)$.
Using the Jackson-Feenberg formula\cite{Jackson1},\cite{Jackson2} and Eq. (\ref{11}) together with hypernetted chain equations (\ref{h7a}) and (\ref{h7}), we can write the energy per particle in the form:

\begin{eqnarray}
\frac{E}{N} &=&\dfrac{1}{N}\langle\psi\vert H \vert \psi\rangle \nonumber \\
&=& E_{F}-\frac{\hbar ^{2}}{4m}\rho _{0}^{2}\int dx\left[ g_{B}(x)-1\right]
\left[ \nabla ^{2}\frac{\sigma ^{2}}{\nu }+2d(x)\nabla ^{2}\sigma (x)\right]
+  \label{g3} \nonumber \\
&&\frac{\rho _{0}}{2}\int dxg(x)\left[ -\frac{\hbar ^{2}}{4m}\nabla ^{2}\ln
f^{2}(x)+V(x)\right].
\label{eq44}
\end{eqnarray}
Replacing $f^{2}(x)$ in the previous equation by its expression given in Eq. (\ref{h5}) we obtain

\begin{equation}
\frac{E}{N}=E_{F}+\varepsilon _{B}+\varepsilon _{S}+\varepsilon _{M},
\label{g4}
\end{equation}
where

\begin{align}
\varepsilon _{B} =\frac{\rho}{2}\int dx\sqrt{g(x)}\left[ -\frac{\hbar \nonumber 
^{2}}{m}\nabla ^{2}+V(x)\right] \sqrt{g(x)}-&\frac{1}{N}\sum_{q}\frac{\hbar 
^{2}q^{2}}{8m}\left( \widetilde{S}(q)-1\right) \nonumber \\
&\left[ \widetilde{E}_{bb}(q)+\frac{\left(\widetilde{S}(q)-1\right) ^{2}}{\widetilde{S}(q)}\right]   
\nonumber 
\end{align}

\begin{eqnarray}
\varepsilon _{S} &\equiv &\rho\int dx\varepsilon _{S}(x)  \nonumber \\
&=&-\frac{\hbar ^{2}\rho _{0}}{8m}\int dx\left\{ \left[ g_{B}(x)-1\right]
\left[ \nabla ^{2}\frac{\sigma ^{2}(x)}{\nu }+2d(x)\nabla ^{2}\sigma
(x)\right] -g(x)\ln \frac{g(x)}{g_{B}(x)}\right\}  \nonumber 
\end{eqnarray}

\begin{eqnarray}
\varepsilon _{M} &\equiv &\frac{1}{N}\sum_{q}\varepsilon _{M}(q)=
-\frac{1}{N}\sum_{q}\frac{\hbar ^{2}q^{2}}{8m}\left( \widetilde{S}(q)-1\right) 
\left\{ \frac{2\widetilde{N}_{ee}(q)+2\widetilde{N}_{be}(q)-\widetilde{N}_{be}^{2}(q)}{1+\widetilde{N}_{ee}(q)}\right.- \nonumber
\end{eqnarray}
\begin{eqnarray}
\hspace{3cm} 2 \left. \widetilde{S}(q)\frac{\left[ \widetilde{N}_{ee}(q)+\widetilde{N}_{be}(q)\right] \left[ 2-\widetilde{N}_{be}(q)+\widetilde{N}_{ee}(q)\right] }{
\left[ 1+\widetilde{N}_{ee}(q)\right] ^{2}}\right\}
\label{g5}
\end{eqnarray}

The first term has the same form as the energy per particle for a system of bosons. It is a functional of the correlation function
$g(x)$ and the structure factor $\widetilde{S}(q)$. The other two terms are specific of fermionic systems  and they depend of functions which can be calculated from Eqs. (\ref{h7a}) and (\ref{h7}).

The energy (\ref{g4}) can be taken as functional of just the correlation function $g(x)$ since the remainder functions
$N_{be}(x)$, $N_{ee}(x)$, $d(x)$, $g_{B}(x)$, etc. can be solved in terms of it by using the hypernetted chain equations (\ref{h7a}) and (\ref{h7}).
Thus $g(x)$ is given by Euler equation: 

\begin{equation}
\frac{\delta (E\left\lbrace g(x)\right\rbrace /N)}{\delta g(x)}=0  \label{g6}
\end{equation}
The presence of the functions $N_{be}(x)$, $N_{ee}(x)$, $d(x)$, $g_{B}(x)$ etc. difficult the evaluation of the functional derivative. In order to simplify the calculation, we follow a scheme originally proposed by Lannto and Siemens\cite{Lantto1}. 
To proceed we consider the energy per particle (Eq. \ref{g4}) as a functional of the seven unknowns: 
$E\left\lbrace g(x),D_{be}(x),D_{ee}(x),d(x),N_{be}(x),N_{ee}(x),n(x)\right\rbrace$. Then we minimize this functional with respect to the seven independent functions with the restriction that hypernetted chain equations (\ref{h7a}) and (\ref{h7}) hold. This is equivalent to reach the extreme of function

\begin{equation}
\Upsilon =E-\sum_{i=1}^{3}\int dx\lambda _{i}(x)\left[
N_{i}(x)-(rhs)_{i}(x)\right] -\sum_{i=1}^{3}\frac{1}{N}\sum_{q}\widetilde{\mu}
_{i}(q)\left[ \widetilde{D}_{i}(q)-\widetilde{(rhs)}_{i}(q)\right]  \label{g7}
\end{equation}
where $\lambda _{i}(x)$ y $\mu _{i}(q)$ are Lagrange multipliers, $N_{i}(x)$ equals $N_{be}(x)$, $N_{ee}(x)$ and $n(x)$ and $D_{i}(x)$ equals $D_{be}(x)$, $D_{ee}(x)$ and $d(x)$  for $i=1,2$ and $3$, respectively. The functions $(rhs)_{i}(x)$ and $\widetilde{(rhs)}_{i}(q)$ with $i=1,2$ and $3$ represent the right hand side of equations (\ref{h7a}) and (\ref{h7}), respectively. 

The function $\Upsilon $ is independently varied with respect to the seven function $g,D_{be},D_{ee},d,N_{be},N_{ee}$ y $n$ and with respect to the Lagrange multipliers too. The variations with respect to the Lagrange multipliers give the hypernetted equations (\ref{h7a}) and (\ref{h7}) again. On the other hand, the variations with respect to $D_{i}(x)$ , say with respect to $D_{be}(x)$, $D_{ee}(x)$ y $d(x)$, yield:
\begin{equation}
\frac{\partial \Upsilon }{\partial D_{i}(x)}=\frac{\partial \varepsilon _{S}}{\partial D_{i}(x)}+\lambda _{i}(x)\frac{\partial N_{i}(x)}{\partial
D_{i}(x)}-\mu _{i}(x)= 0  \label{g8}
\end{equation}

In Eq. (\ref{g8}), the derivatives $\frac{\partial N_{i}(x)}{\partial D_{i}(x)}$ are calculated using the right hand side of 
Eqs. (\ref{h7}) and $\mu _{i}(x)$ is the inverse Fourier transform of $\widetilde{\mu} _{i}(q)$. 
The variations with respect to the functions $\widetilde{N}_{i}(q)$, say, $\widetilde{N}_{be}(q)$, $\widetilde{N}_{ee}(q)$
and $\widetilde{n}(q)$ give:

\begin{equation}
\frac{\partial \Upsilon }{\partial \widetilde{N}_{i}(q)}=\frac{\partial \widetilde{\varepsilon}
_{M}(q)}{\partial \widetilde{N}_{i}(q)}+\widetilde{\mu} _{i}(q)\frac{\partial \widetilde{D}_{i}(q)}{\partial
\widetilde{N}_{i}(q)}-\widetilde{\lambda} _{i}(q) = 0  \label{g9}
\end{equation}

The derivatives $\frac{\partial \widetilde{D}_{i}(q)}{\partial \widetilde{N}_{i}(q)}$ are calculated by using
the right hand side of Equations (\ref{h7a}) and $\widetilde{\lambda} _{i}(q)$ is the
Fourier transform of $\lambda _{i}(x)$. Eqs. (\ref{g8}) y (\ref{g9}) are a set of six linear equations for the Lagrange multipliers 
$\lambda _{i}(x)$ y $\mu _{i}(x)$. The fact that we have a linear system is what makes Lannto and Siemens approach useful\cite{Lantto1}. 

Finally, variation of the functional $\Upsilon $ with respect to $\sqrt{g(x)}$ 
yields the zero energy scattering equation:

\begin{equation}
\left[ -\frac{\hbar }{m}\nabla ^{2}+V(x)+w_{B}(x)+w_{S}(x)+w_{M}(x)\right] \sqrt{g(x)}=0  
\label{g10}
\end{equation}
Here the induced potential $w_{B}$ is the same as for bosons:

\begin{equation}
\widetilde{w}_{B}(q)=\frac{\hbar ^{2}q^{2}}{4m}\left\{ \widetilde{E}_{bb}(q)+
\left[ \widetilde{S}(q)-1\right] \frac{\partial \widetilde{E }_{bb}(q)}{\partial \widetilde{S}(q)}+
\frac{\left[\widetilde{S}(q)-1\right] ^{2}\left[ 2\widetilde{S}(q)+1\right] }{\widetilde{S}^{2}(q)}\right\} .
\label{34}
\end{equation}
For fermions we have the additional induced potentials $w_{S}(x)$ and $w_{M}(x)$:

\begin{eqnarray}
w_{S}(x) &=&2\frac{\partial \varepsilon _{S}(x)}{\partial g(x)}%
+2\sum_{i=1}^{3}\lambda _{i}(x)\frac{\partial N_{i}(x)}{\partial g(x)}
 \nonumber \\
\widetilde{w}_{M}(q) &=&2\frac{\partial \widetilde{\varepsilon}_{M}(q)}{\partial \widetilde{S}(q)}
+2\sum_{i=1}^{3}\widetilde{\mu}_{i}(q)\frac{\partial \widetilde{D}_{i}(q)}{\partial \widetilde{S}(q)}.
\label{g11}
\end{eqnarray}

The derivatives $\frac{\partial \varepsilon _{S}(x)}{\partial g(x)}$ and 
$\frac{\partial \widetilde{\varepsilon}_{M}(q)}{\partial \widetilde{S}(q)}$ are evaluated from
Eqs. (\ref{g5}), whereas $\frac{\partial N_{i}(x)}{\partial g(x)}$ and 
$\frac{\partial \widetilde{D}_{i}(q)}{\partial \widetilde{S}(q)}$ are calculated by using the right hand side
of Eqs. (\ref{h7a}) and (\ref{h7}), respectively.

The procedure to solve the system of equations (\ref{h7a}), (\ref{h7}), (\ref{g8}), (\ref{g9}) and (\ref{g10}) with Eqs. (\ref{34}) and (\ref{g11}) is a follows. Firstly we solve Eq. (\ref{g10}) as for bosons (with $w_{S}(x)=w_{M}(x)=0$ and $w_{B}(x)$ the inverse Fourier transform of Eq.(\ref{34})). The correlation function $g(x)$ so calculated is used in the hypernetted chain equations (\ref{h7a}) and (\ref{h7}) to obtain by iteration the six functions $D_{i}$ and $N_{i}$ $(i = 1,2,3)$. With these functions we solve the linear system of equations (\ref{g8}-\ref{g9}) for $\lambda_{i}$ and $\mu_{i}$ that allow us to construct the potentials $w_{S}(x)$ and $w_{M}(x)$ (Eq. \ref{g11}). With these potentials we solve again Eq. (\ref{g10}) and the iteration is continued until selfconsistence is attained. In the examples below few iterations (4 or 5) are enough to achieve convergence.

\subsection{Elementary diagrams}
\begin{figure}[here!]
 \centering
 \includegraphics[width=0.6\textwidth]{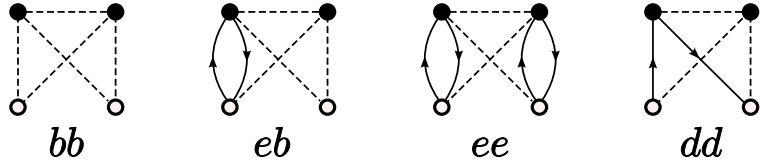}
 \caption{Diagrams used as generators of the elementary graphs. }
\label{fig1}
\end{figure}

\begin{figure}[here!]
 \centering
 \includegraphics[width=0.7\textwidth]{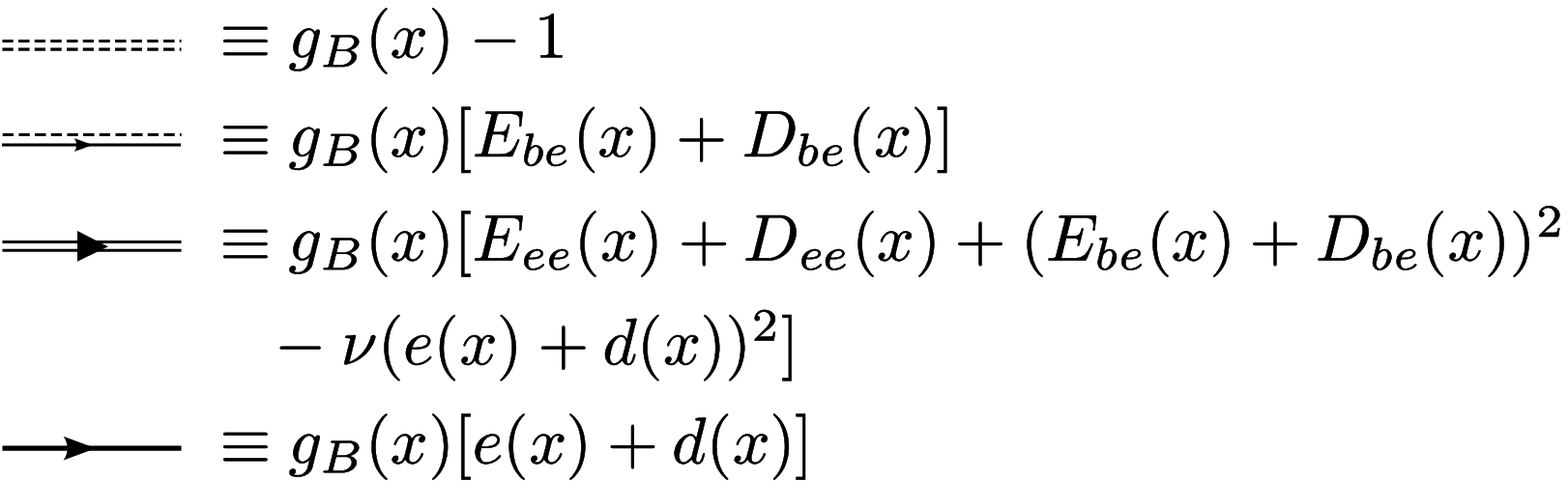}
 \caption{Generalized bonds.}
\label{fig2}
\end{figure}

Up to now the matrix $E$, whose elements are the sum of the different types of elementary diagrams, is taken as known. We have already commented about the difficulties involved in its calculation which makes this an almost impossible task, so that some kind of approximation, that implies to avoid many of the graphs, is necessary. According to the order of retained elementary diagrams  we have classified the resulting approximation as QHNC/n, where $n$ denotes the maximum order. In particular if we ignore all the elementary graphs we have the QHNC/0 approximation, which in the literature is often simply called HNC. In the calculations below, we will take $n=4$ that corresponds to the smallest (non-null) order. Even for a given order, to sum the whole set of diagrams is very hard, so in our calculations we apply the following strategy. For each kind of graph ($bb, eb, ee$ and $dd$) we take as generators the ones shown in Fig. \ref{fig1} and observe that replacing some of their bonds by some of the generalized bonds defined in Fig \ref{fig2} we recover many of the elementary graphs of the same kind and order as the generator one.

\begin{figure}[here!]
 \centering
 \includegraphics[width=0.6\textwidth]{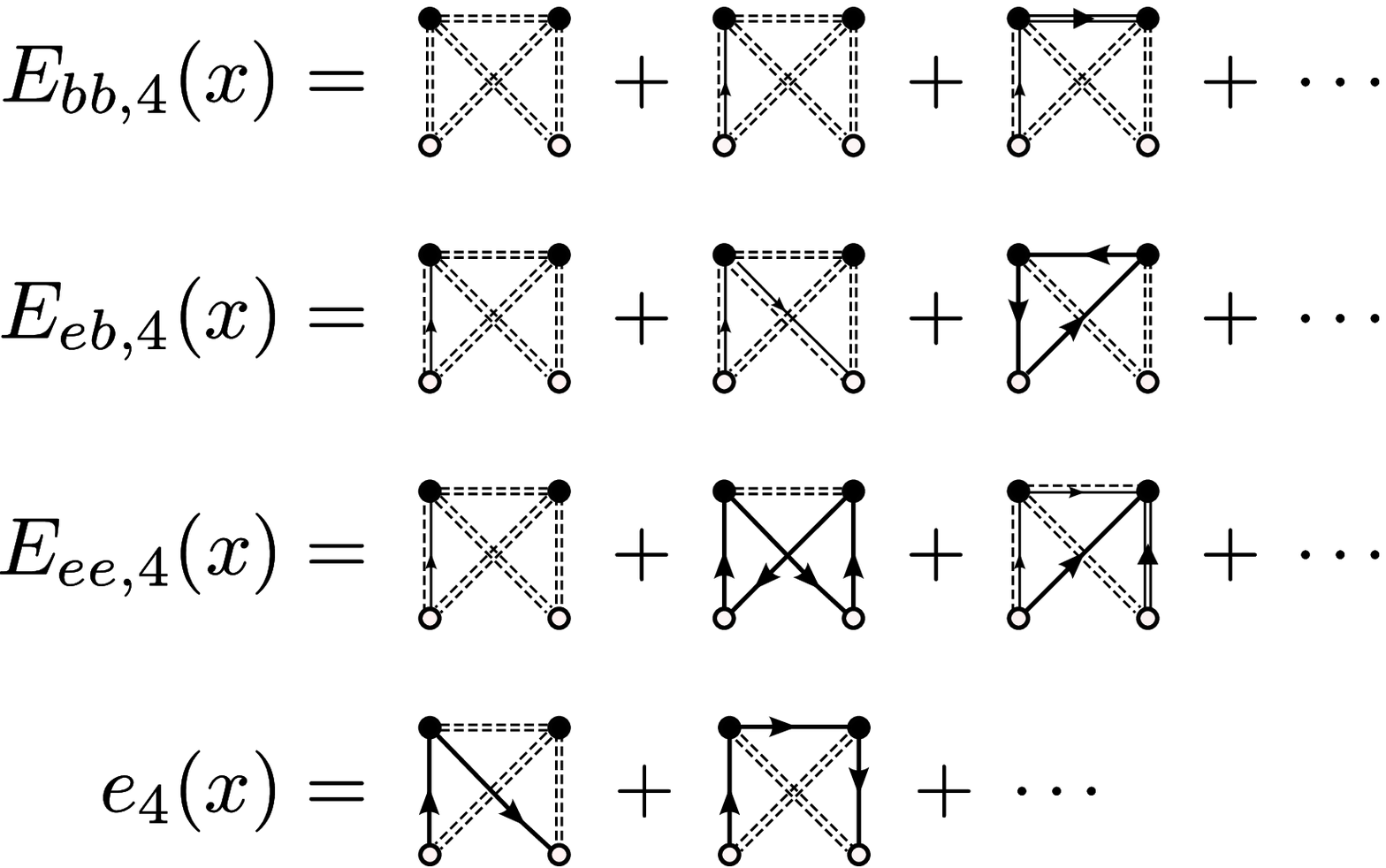}
 \caption{Generalized elementary diagrams used in our calculations}
\label{fig3}
\end{figure}

Specifically in our calculations in next section we consider the diagrams shown in Fig.\ref{fig3}

\subsection{QHNC for binary mixtures}
\label{QHNC_mixtures}

Since several of the applications in Section III involve two component systems, we outline here the QHNC equations for binary mixtures. The supra-indices $i,j$, $(i,j=1,2)$ denote the species in the mixture. 

The energy equation (Eq. \ref{eq44}) now reads  
\begin{eqnarray}
\frac{E}{N} =\sum_{i}\frac{\hbar^{2}}{4m^{i}}(\rho^{i})^{2}\int dx\left[ g_{B}^{ii}(x)-1\right]
\left[ \nabla ^{2}\frac{(\sigma^{i})^{2}}{\nu^{i} }+2d^{i}(x)\nabla ^{2}\sigma^{i} (x)\right]
+  \nonumber \\
\sum_{i,j}\frac{1}{2}\rho^{i}\rho^{j}\int dxg^{ij}(x)\left\lbrace -\frac{\hbar ^{2}}{4\mu^{ij}}\nabla ^{2}\left[ \ln g_{B}^{ii} (x)-
D_{bb}^{ij}(x)\right] +V^{ij}(x)\right\rbrace 
\end{eqnarray}
with $\mu^{ij}$ the reduced mass.
The QHNC equations \ref{h7a} and \ref{h7} yield
\begin{eqnarray}
\widetilde{D}_{be}(q) &=&(I+\widetilde{N}_{be}(q))(I-\widetilde{N}_{be})^{T}\widetilde{S}(q)\left[I-(I-\widetilde{N}_{be})   \nonumber
                          (I-\widetilde{N}_{ee})^{-1}\right] - \nonumber  \\
                       &  & -\widetilde{N}_{be}-I+(I-\widetilde{N}_{ee})^{-1} \nonumber \hspace*{10 cm}  
\nonumber \\
\widetilde{D}_{ee}(q) &=& I-\widetilde{N}_{ee}-(I+N_{ee})^{-1}+\widetilde{S}(q)
\left[I-(I-\widetilde{N}_{be})(I+\widetilde{N}_{ee})^{-1}\right]  \hspace*{10 cm}  \nonumber \\
&-&(I+\widetilde{N}_{ee})^{-1}(I-\widetilde{N}_{be})^{T}\widetilde{S}(q)+(I+\widetilde{N}_{ee})^{-1}(I-\widetilde{N}_{be})^{T}\widetilde{S}(q)(I-\widetilde{N}_{be})(I+\widetilde{N}_{ee})^{-1} \nonumber \\
\nonumber \\
\widetilde{d}^{i}(q) &=&\left\{ 
\begin{array}{cc}
-\widetilde{n}^{i}(q) & \text{para }\left| q\right| <k_{F} \nonumber \\ 
-\frac{(\widetilde{n}^{i})^{2}(q)}{1+\widetilde{n}^{i}(q)} & \text{para }\left| q\right| >k_{F}
\end{array}
\right.     
\end{eqnarray}
\begin{equation}
\end{equation}
and
\begin{eqnarray}
N_{be}^{ij}(x) &=& g_{B}^{ij}(x)\left[D_{be}^{ij}(x)\right] -D_{be}^{ij}(x)  \nonumber  \\
N_{ee}^{ij}(x) &=& g^{ij}(x)-D_{ee}^{ij}(x)  \nonumber \\
n^{i}(x) &=& g_{B}^{ii}(x)\left[ e^{i}(x)+d^{i}(x)+\frac{\sigma^{i} (x)}{\nu ^{i}}\right] -d^{i}(x)-\frac{\sigma^{i} (x)}{\nu^{i} }  \nonumber \\
g^{ij}(x) &=& g_{B}^{ij}(x)\left\{E_{ee}^{ij}(x)+D_{ee}^{ij}(x)+\left[E_{be}^{ij}+D_{be}^{ij}(x)\right] ^{2}-\nu^{i}\delta_{ij}(e^{i}(x)+d^{i}(x)+\sigma^{i}/\nu^{i})^{2}\right., \nonumber 
\end{eqnarray}
\begin{equation}
\end{equation}
respectively. The symbols $N$ and $D$ denote matrices whose elements correspond to species pairs.  

The function whose extreme must be found is

\begin{equation}
\Upsilon =E-\sum_{i=1}^{3}\sum_{j,k}\int dx\lambda _{i}^{jk}(x)\left[N_{i}^{jk}(x)-(rhs)_{i}^{jk}(x)\right] - \nonumber
\end{equation}
\begin{equation}
  -\sum_{i=1}^{3}\sum_{j,k}\frac{1}{\sqrt{N_{j}N_{k}}}\sum_{q}\widetilde{\mu}
_{i}^{jk}(q)\left[ \widetilde{D}_{i}^{jk}(q)-\widetilde{(rhs)}_{i}^{jk}(q)\right],
\end{equation}
being

\begin{equation}
\frac{\partial \Upsilon }{\partial D_{i}^{jk}(x)}=\frac{\partial E}{\partial D_{i}^{jk}(x)}+\lambda _{i}^{jk}(x)\frac{\partial N_{i}^{jk}(x)}{\partial D_{i}^{jk}(x)}-\mu _{i}^{jk}(x)= 0  
\end{equation}

and

\begin{equation}
\frac{\partial \Upsilon }{\partial \widetilde{N}_{i}^{jk}(q)}=
\frac{\partial E}{\partial \widetilde{N}_{i}^{jk}(q)}+\widetilde{\mu} _{i}^{jk}(q)\frac{\partial \widetilde{D}_{i}^{jk}(q)}{\partial
\widetilde{N}_{i}^{jk}(q)}-\widetilde{\lambda} _{i}^{jk}(q) = 0  
\end{equation}
the extremal equations. Again, variations of the functional $\Upsilon $ with respect to $\sqrt{g^{ij}(x)}$ 
yields:

\begin{equation}
\left[ -\frac{\hbar }{2\mu^{ij}}\nabla ^{2}+V^{ij}(x)+w_{B}^{ij}(x)+w_{S}^{ij}(x)+w_{M}^{ij}(x)\right] \sqrt{g^{ij}(x)}=0  
\end{equation}
where functions $w_{B}^{ij}(x), w_{S}^{ij}(x)$ and $w_{M}^{ij}(x)$ are given by

\begin{equation}
 w_{B}^{ii}(q)=-\dfrac{\hslash^{2}q^{2}}{4\rho^{i}}\left\lbrace \dfrac{2\widetilde{S}^{ii}(q)-3}{m^{i}}+
\dfrac{\widetilde{S}^{jj}(q)^{2}/m^{i}+\widetilde{S}^{ij}(q)^{2}/m^{j}}2
{(\widetilde{S}^{11}(q)\widetilde{S}^{22}(q)-\widetilde{S}^{12}(q)^{2})^{2}}\right\rbrace 
\end{equation}

\begin{equation}
 w_{B}^{ij}(q)=-\dfrac{\hslash^{2}q^{2}}{4\sqrt{\rho^{i} \rho^{j}}}\left\lbrace \dfrac{\widetilde{S}^{ij}(q)}{\mu^{ij}}-
\dfrac{(\widetilde{S}^{jj}(q)/m^{i}+\widetilde{S}^{ii}(q)/m^{j})\widetilde{S}^{ij}(q)}
{(\widetilde{S}^{11}(q)\widetilde{S}^{22}(q)-\widetilde{S}^{12}(q)^{2})^{2}}\right\rbrace 
\end{equation}

\begin{eqnarray}
w_{S}^{jk}(x) &=&2\frac{\partial E}{\partial g^{jk}(x)}%
+2\sum_{i}\lambda _{i}^{jk}(x)\frac{\partial N_{i}^{jk}(x)}{\partial g^{jk}(x)}
 \nonumber \\
\widetilde{w}_{M}^{jk}(q) &=&2\frac{\partial E}{\partial \widetilde{S}^{jk}(q)}
+2\sum_{i}\widetilde{\mu}_{i}^{jk}(q)\frac{\partial \widetilde{D}_{i}^{jk}(q)}{\partial \widetilde{S}^{jk}(q)}.
\end{eqnarray}

It should be mentioned that a somewhat different version of QHNC for multicomponent systems in 3D has been reported by Lantto \cite{Lantto3}.

\section{Applications to quantum wires}
Given a heterostructure made of several layers of suitable semiconductor materials,
by means of diverse experimental techniques it has been possible to conveniently modify the band structure of some of the layers \cite{Reed1},\cite{Butcher1}. 
\begin{figure}[here]
\centering
 \includegraphics[width=0.65\textwidth]{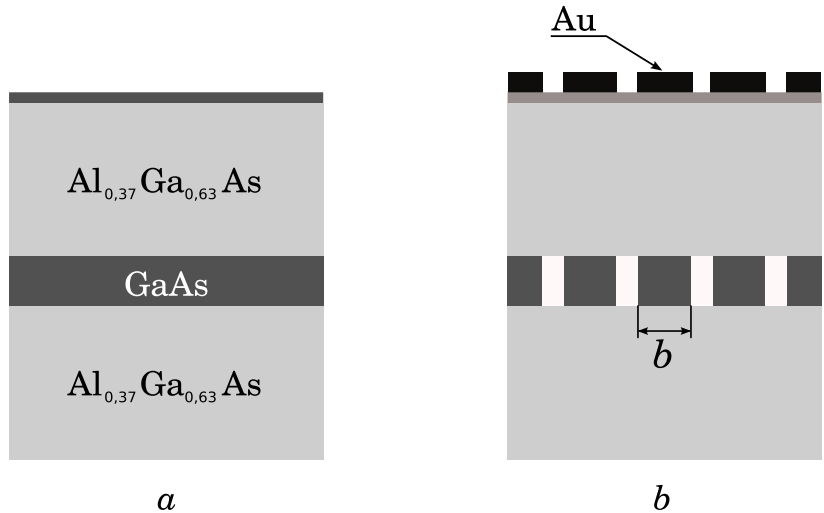}
\caption{Schematic construction of quantum wires. \textit{a}) quantum well: the carriers are constrained to move in the layer of GaAs. \textit{b}) Ga ions, which are implanted into the GaAs layer after being collimated by a gold mask, modify the layer band structure in such a way that the carriers are constrained to move along the quantum wires of width \textit{b}.}
\label{fig: cables}
\end{figure}
The result is that movement of carriers is restricted in one, two or even the three possible directions. In the first case they are confined to a slab of a few nanometers width that can be described, within Sommerfeld-Pauli picture, as a quasi-bidimensional system of fermions known in the literature as ``quantum well''(see Fig. \ref{fig: cables} \textit{a})\cite{Dingle1}. 
If the carriers movement in this quasi-bidimensional structure is even resticted in a second direction to a width $b$ then we have the quasi-onedimensional structures we are interested in: the quantum wires\cite{Sakai1},\cite{Petroff1} (see Fig. \ref{fig: cables} \textit{b}). 
Typical values  of $b$ run from tenths to hundreds of nanometers.

\begin{figure}[here]
\centering
 \includegraphics[width=0.65\textwidth]{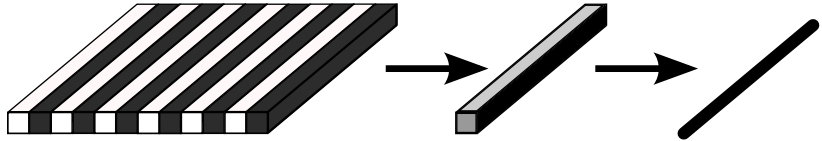}
\caption{Assuming that the various parallel wires are not coupled, we can isolate one of these quasi-unidimensional devices and treat it as a rigorously one dimensional system by using effective pair interaction potentials}
\label{cc1d}
\end{figure}

Within this context, our model of quantum wire will be, according to which properties one wishes to account for, or a one component system of charged fermions (electrons) or a binary mixture whose particles have opposite sign (electrons and holes). These particles can move along a quasi-one dimensional region which is practically unlimited in one of the directions but transversally very narrow.
Here we are thinking in just one wire. However in the experimental array there are several wires arranged each one practically parallel to the others.
In principle it has sense to study an isolated wire ( Fig. \ref{cc1d}) if we assume that the interactions among the carriers belonging to different wires are negligible or if we incorporate in some way these interactions as a sort of external field acting on the particles of the isolated wire. The effect of coupling between two wires is explicitly considered in Subsection C. 

Besides, although in principle we can think of an isolated wire as a quasi-one dimensional region, actually we wish to describe it from a \textit{rigorously one dimensional} point of view (Fig. \ref{cc1d}). This can be achieved by defining appropriate effective pair potentials.

In this Section we apply the hypernetted chain equations obtained above to describe some properties of quantum wires. In the next Subsection (A) we consider some aspects related with their conductor nature, such as the Wigner crystallization. To this we model an isolated quantum wire, within Sommerfeld-Pauli picture, as a one dimensional electron gas. In Subsection B we study properties that are related with the semiconductor behavior instead, particularly the photoluminescence phenomenon. In this case the quantum wire is seen as a 1D mixture of electrons and holes. Finally in Subsection C we consider a pair of coupled quantum wires in order to study how the correlations of carriers in one quantum wire is affected by the presence of a second one.

\subsection{Quantum wires as conductors}
\label{cce}
In this Subsection quantum wires are modeled as a one dimensional electron gas, say a rigorously one dimensional system in which the $N$ electrons interact via long range pair potentials and freely move along the $x$ axis. 
For this system we calculate pair correlation functions and structure factors in the QHNC approximation and also perform variational quantum Monte Carlo (VQMC) in  simulations in order to compare our hypernetted chain results.

\subsubsection{Effective pair potentials}

We consider the effective mass and envelope function approximations and give account of the transversal confinement of wires carriers through a parabolic one particle potential of the form:  

\begin{equation}
U(y_{1},z_{1})=\frac{\hbar ^{2}\left[ y_{1}^{2}+z_{1}^{2}\right] }{8m^{*}b^{4}}  \label{cc1}
\end{equation}
with $m^{*}$ the electron effective mass and $b$ the wire width. 

For this potential the transversal one particle wave functions are those of an harmonic oscillator. If the separation between the energy levels is large enough then we can assume that the electrons transversally are in the ground state with wave function

\begin{equation}
\phi (y_{1},z_{1})=\frac{1}{\sqrt{2\pi }b}\exp (-\frac{y_{1}^{2}+z_{1}^{2} }{4b^{2}})
\label{cc2}
\end{equation}
so that the longitudinal envelope function $\psi(x)$ where $x=x_{1}-x_{2}$ must verify the one dimensional effective Hamiltonian
\begin{equation}
H=-\frac{1}{r_{s}^{2}}\frac{d^{2}}{dx^{2}}+\frac{2}{r_{s}}V_{eff}(x)  \label{cc3}
\end{equation}
where $V_{eff}$ is the effective pair potential

\begin{equation}
V_{eff}(x)=-\int \frac{e^{2}|\phi(y_{1},z_{1}|^{2}|\phi(y_{2},z_{2}|^{2}dy_{1}dz_{1}dy_{2}dz_{2}} {\varepsilon_{0}\sqrt{(x^{2}+(y_{1}-y_{2})^{2}+(z_{1}-z_{2})^{2}}} \\ \nonumber
\end{equation}
\begin{equation}
= e^{2}\frac{\sqrt{\pi}}{2b} \exp (\frac{(x^{2}}{4b^{2}}) \text{erfc} (\frac{\vert x \vert }{2b})
\label{cc4}
\end{equation}
with $e$ being the charge of an electron.

We see that all the system properties depend on just two parameters: $b$ and $r_{s}$. The parameter $r_{s}=(2\rho a_{B}^{*})^{-1}$ (with $a_{B}=\hbar ^{2}\epsilon_{0} /m^{*}e^{2}$ the Bohr radius for a medium of dielectric constant $\epsilon_{0}$) is a measure of the mean distance between two electrons. It is inversely related to the density and also measures the system coupling as expressed by the ratio between the potential and kinetic energies.

Eventually we get for $ V_{eff}(x)$ another expression which also captures the essential of the interactions between confined electrons and is very frequently used in the literature, the so called Schultz potential\cite{Schulz1}:
\begin{equation}
\label{schultz} 
 V_{eff}(x)=\dfrac{e^{2}}{\sqrt{x^{2}+b^{2}}}. 
\end{equation} 

\subsubsection{Pair correlation functions}
In the form we have presented the QHNC equations in Section II, the pair correlation functions, as defined by Eqs.\ref{12} and \ref{8}, are proportional to the probability density of finding an electron at a distance $x$ from another one irrespective of their spins. In this subsection we compare our QHNC correlation functions with those obtained from VQMC simulations. Since is relatively simple to explicitly incorporate the spins in the simulations, we additionally show the spin dependent pair correlation functions calculated with VQMC so helping to better understand the system structural behavior. 
To define this last correlation functions let consider a system of $N=N_{\uparrow}+N_{\downarrow}$ ($N_{\uparrow}$, $N_{\downarrow}$ number of electrons with spin up and spin down, respectively) moving along a segment of length $L$. Then

\begin{equation}
g_{ss^{\prime }}(x)=\frac{1}{\rho _{s}\rho _{s^{\prime }}L}\sum_{i\neq
j}\left\langle \delta (x_{i}^{s^{\prime }}-x_{i}^{s}-x\right\rangle
\label{corre1}
\end{equation}
where $\rho _{s}= N_{s}/L$ is the mean density for electrons with spin $s$ ($s=\uparrow$ or $\downarrow$).

These functions describe the probability density of finding an electron with spin $s^{\prime }$ 
separated a distance $x$ from another electron with spin $s$. This way the correlations between electrons with parallel and antiparallel spins are distinguished.

The correlation functions for polarized and non-polarized phases read

\begin{eqnarray}
&&\left. 
\begin{array}{c}
g_{\uparrow \uparrow }(x)=\frac{4L}{N^{2}}\sum_{i\neq j}\left\langle \delta
(x_{i}^{\uparrow }-x_{j}^{\uparrow }-x)\right\rangle \\ \nonumber
g_{\uparrow \downarrow }(x)=\frac{4L}{N^{2}}\sum_{i\neq j}\left\langle
\delta (x_{i}^{\uparrow }-x_{j}^{\downarrow }-x)\right\rangle
\end{array}
\right\} \text{non-polarized}  \label{corre2} \\ \nonumber
&& 
\begin{array}{c}
\left. g_{\uparrow \uparrow }(x)=\frac{L}{N^{2}}\sum_{i\neq j}\left\langle
\delta (x_{i}^{\uparrow }-x_{j}^{\uparrow }-x)\right\rangle \right\}
\end{array}
\text{fully polarized}  
\end{eqnarray}

We can combine these functions to define the so called numeric and magnetic correlation functions:

\begin{eqnarray}
g(x)= g_{nn}(x) &=&\frac{1}{2}\left[ g_{\uparrow \uparrow }(x)+g_{\uparrow
\downarrow }(x)\right]  \label{corre3} \\
g_{mm}(x) &=&\frac{1}{2}\left[ g_{\uparrow \uparrow }(x)-g_{\uparrow
\downarrow }(x)\right],  \nonumber
\end{eqnarray}
for non-polarized systems and

\begin{equation}
g_{nn}(x)=g_{mm}(x)=g_{\uparrow \uparrow }(x),  \label{corre4}
\end{equation}
for fully polarized systems.

\begin{figure}[here]
 \centering
 \includegraphics[width=1.0\textwidth]{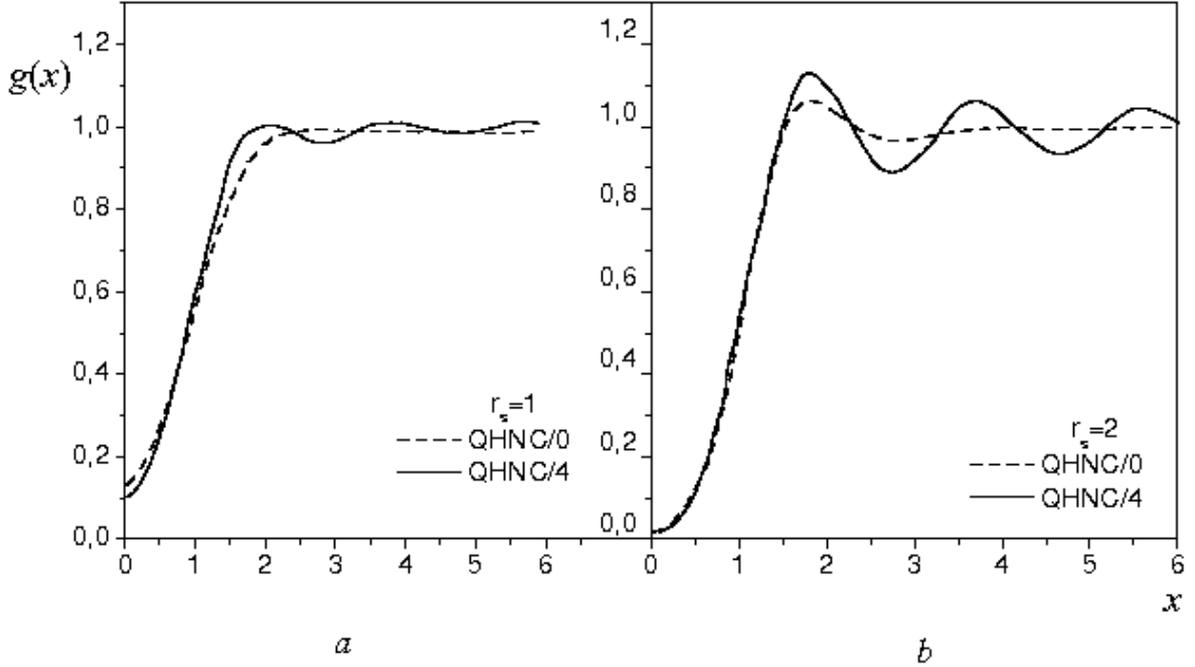}
  \caption{Comparison of QHNC/0 and QHNC/4 versions for $b$ = 0.1 and \textit{a}) $r_{s}=1$; \textit{b}) $r_{s}=2$. The distance $x$ is scaled by the Bohr radius. The pair potential between the electrons is given by Eq. (\ref{cc4}). }
 \label{qhnc04}
\end{figure}

Next we show the correlation functions that we have calculated for the effective pair potential given by Eq. (\ref{cc4}). 
With respect to the QHNC approximation recall that we have mentioned two versions: one in which all the elementary diagrams are ignored. We call it QHNC/0. In the other one, denoted QHNC/4, we include the first elementary graphs of order 4.
In fig \ref{qhnc04} these two versions are compared. 

We observe that the inclusion of the elementary diagrams of order 4 becomes the correlation function more structured. From now on we will take as approximation QHNC the version QHNC/4

As was mentioned, in order to appreciate the goodness of this approximation we compare with the corresponding results obtained from VQMC simulations that, in this way, acts as a virtual laboratory.

\begin{figure}[here!]
 \centering
 \includegraphics[width=1.0\textwidth]{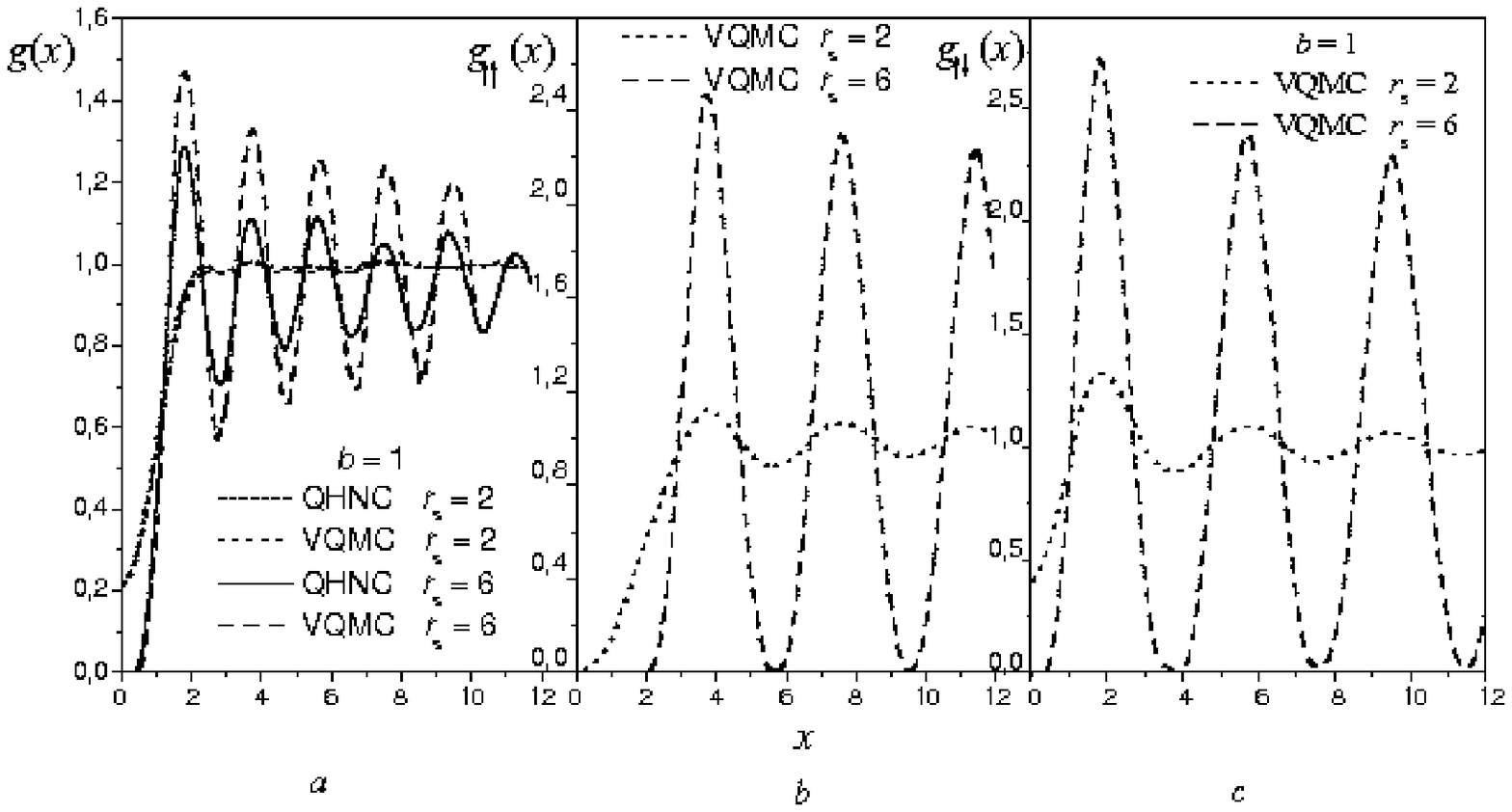}
 \caption{Correlation functions of a 1D system of electrons interacting via the pair potential given by Eq. (\ref{cc4}) for $b=1$ and diverse values of $r_{s}$: \textit{a}) averaged spins; \textit{b})parallel spins; \textit{c}) antiparallel spins.}
 \label{qhnc-vqmc-b1}
\end{figure}
\begin{figure}[here!]
 \centering
 \includegraphics[width=1.0\textwidth]{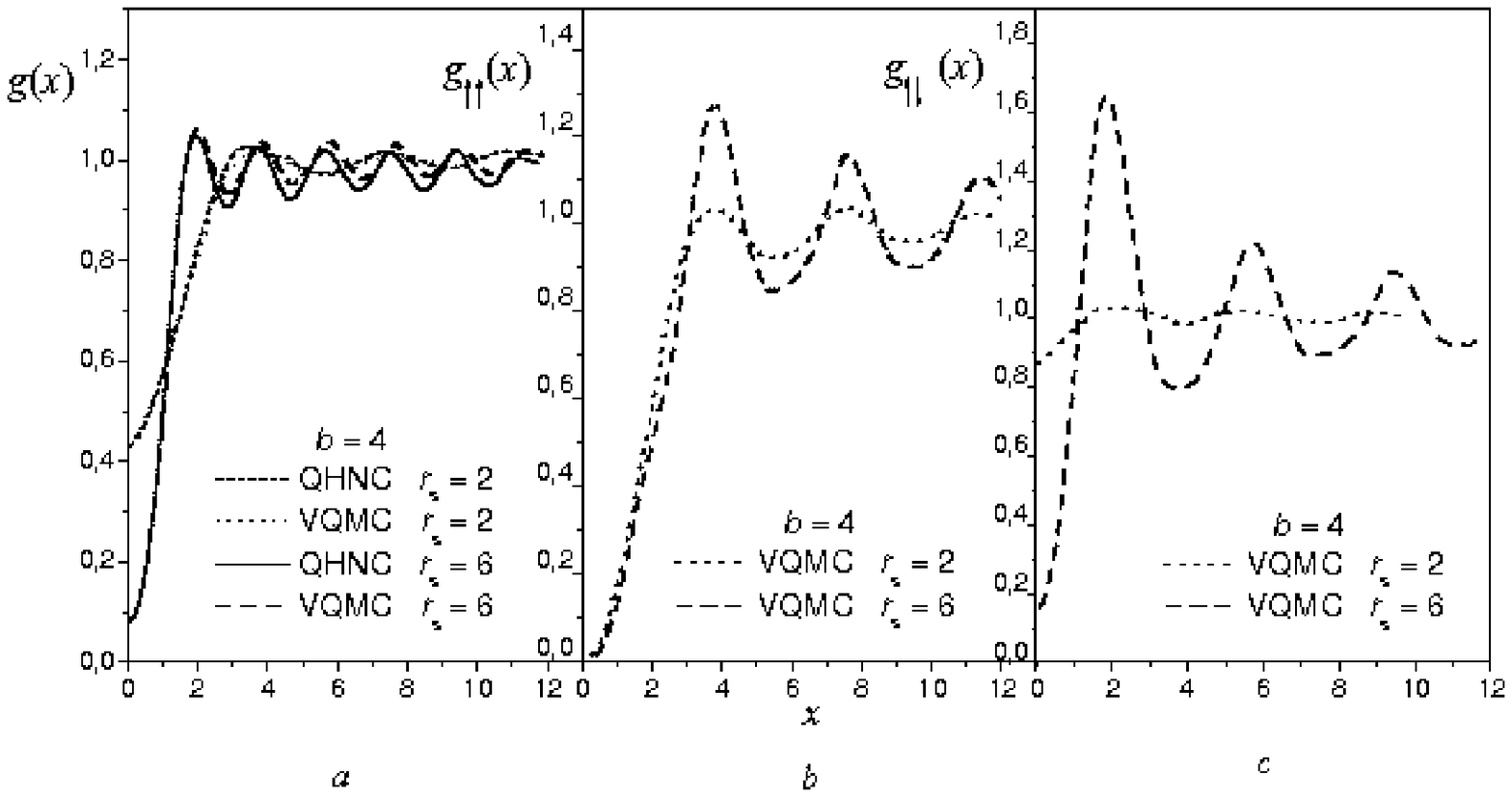}
 \caption{Correlation functions of a 1D system of electrons interacting via the pair potential given by Eq. (\ref{cc4}) for $b=4$ and diverse values of $r_{s}$: \textit{a}) averaged spins; \textit{b})parallel spins; \textit{c}) antiparallel spins.}
 \label{qhnc-vqmc-b4}
\end{figure}

In figures \ref{qhnc-vqmc-b1} \textit{a} and \ref{qhnc-vqmc-b4}
\textit{a}, we show the results obtained from the QHNC theory and VQMC simulations for the correlation function $g(x)$ corresponding to the wire widths $b = 1$ and $4$ (in units of the Bohr radius). We observe that the smaller is the density more structured the system is. This effect is more noticeable for small widths. This is reasonable since when $b$ increases the system behavior tends to the one of a bidimensional system. The larger confinement imposed to the electrons in the case $b=1$ manifests into a larger Pauli repulsion which can clearly be appreciated by comparing the contact values ($x=0$).

The correlation functions $g(x)$ considered in these figures correspond to non polarized paramagnetic electron gas. It should be interesting to analyze the correlations between pairs of particles with parallel and antiparallel spins. The panel \textit{b} of figures \ref{qhnc-vqmc-b1} and \ref{qhnc-vqmc-b4}
show $g_{\uparrow\uparrow}(x)$, whereas that in the corresponding panel \textit{c} we see $g_{\uparrow \downarrow}(x)$. These figures were obtained just from Monte Carlo simulations because the QHNC approximation, as it is considered here, can not allow, in principle, to calculate separately the correlations between electrons with parallel spins and those corresponding to electrons with antiparallel spins. In the graphics the Pauli exclusion hole for $g_{\uparrow\uparrow}(x)$ is apparent. Besides, from the observation of figures \textit{b} and \textit{c}, we conclude that the electrons arrange in such a way that spins up alternate with spins down. When the density grows (say, when degeneracy grows) the curves $g_{\uparrow\downarrow}(x)$ show that the system quickly tends to behave as an ideal gas. Obviously, the main difference with $g_{\uparrow\uparrow}(x)$ is near the origin where the effects of the fermion statistic are more important.

\begin{figure}[here!]
 \centering
 \includegraphics[width=1.0\textwidth]{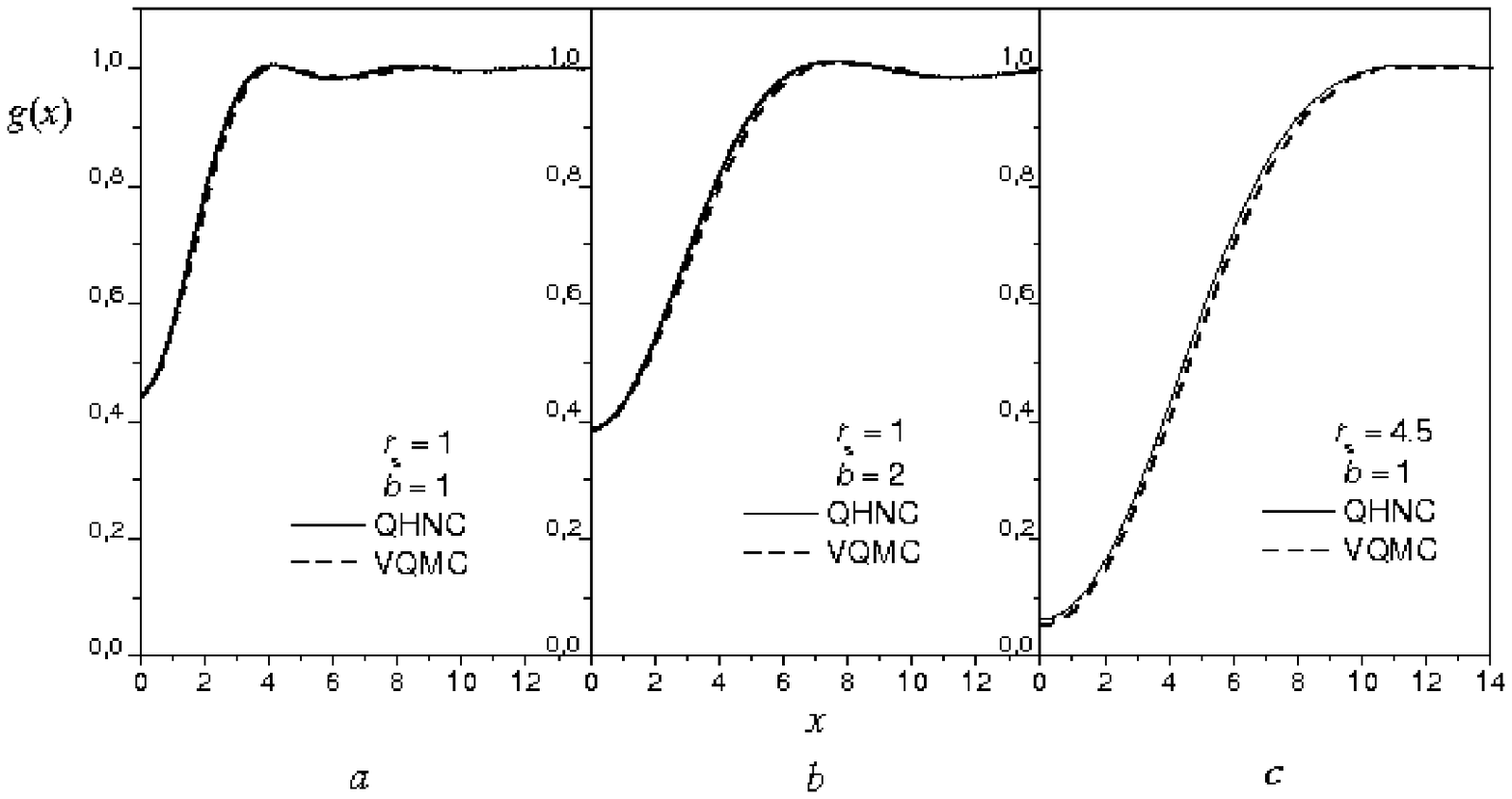}
 \caption{Correlation functions of a 1D system of electrons interacting via the pair potential given by Eq. (\ref{schultz}) for diverse values of $r_{s}$ and $b$: \textit{a}) $r_{s}=1$, $b=1$; \textit{b}) $r_{s}=1$, $b=2$; \textit{c}) $r_{s}=4.5$, $b=1$.}
 \label{qhnc-vqmc-sch}
\end{figure}

We have also have calculated the correlations functions for Schultz potential (Eq.\ref{schultz})  In figure \ref{qhnc-vqmc-sch} we show the results for diverse values of $r_{s}$ and $b$. It is worth mentioning the good agreement of the QHNC and VQMC calculations, better than that achieved with the potential given by Eq. (\ref{cc4}) 

\subsubsection{Structure factors and crystallization}
The correlation functions are adequate to describe the structure of the electron gas in real space, but essentially are theoretic objects. More directly related to X ray or neutron diffraction experiments are the structure factors. These functions are obtained from the intensity of the diffracted radiation (previous elimination of the atomic form factor) and simply relate to the pair correlation functions via Fourier transforms

\begin{equation}
 g_{ss'}(x)= 1+\dfrac{1}{\sqrt{\rho_{s} \rho_{s'}}} \int \dfrac{dk}{2 \pi}e^{ikx} \left[ S_{ss'}(q)-\delta_{ss'}\right],
\label{fe3}
\end{equation}
where the subindices denote the spin of the involved particles.

From the partial structure factors $S_{ss'}$ we define numeric and magnetic structure factors which are linear combinations of the spin-up/spin-up and the spin-up/spin-down structure factors:
\begin{eqnarray}
S_{nn}(k) &=&S_{\uparrow \uparrow }(k)+S_{\uparrow \downarrow }(k)
\label{fe4} \\
S_{mm}(k) &=&S_{\uparrow \uparrow }(k)-S_{\uparrow \downarrow }(k)  \nonumber
\end{eqnarray}

\begin{figure}
 \centering
 \includegraphics[width=1.0\textwidth]{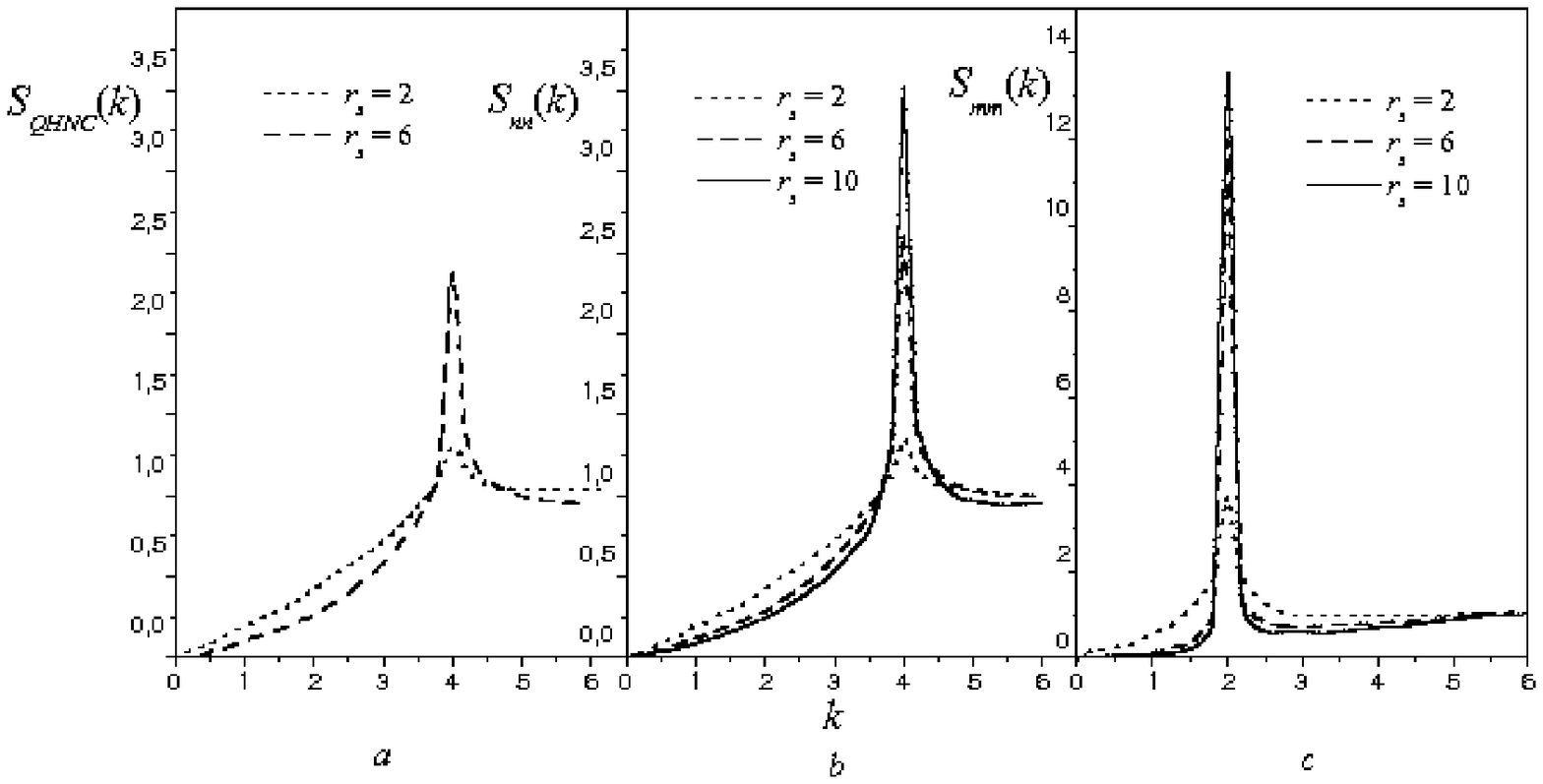}
 \caption{Structure factors for $b$ = 0.1 and diverse values of $r_{s}$. \textit{a}): QHNC; \textit{b}): VQMC numeric ;\textit{c}): VQMC magnetic. Momenta are reduced by the Fermi momentum.}
 \label{skb01}
\vspace{2cm}
\centering
 \includegraphics[width=1.0\textwidth]{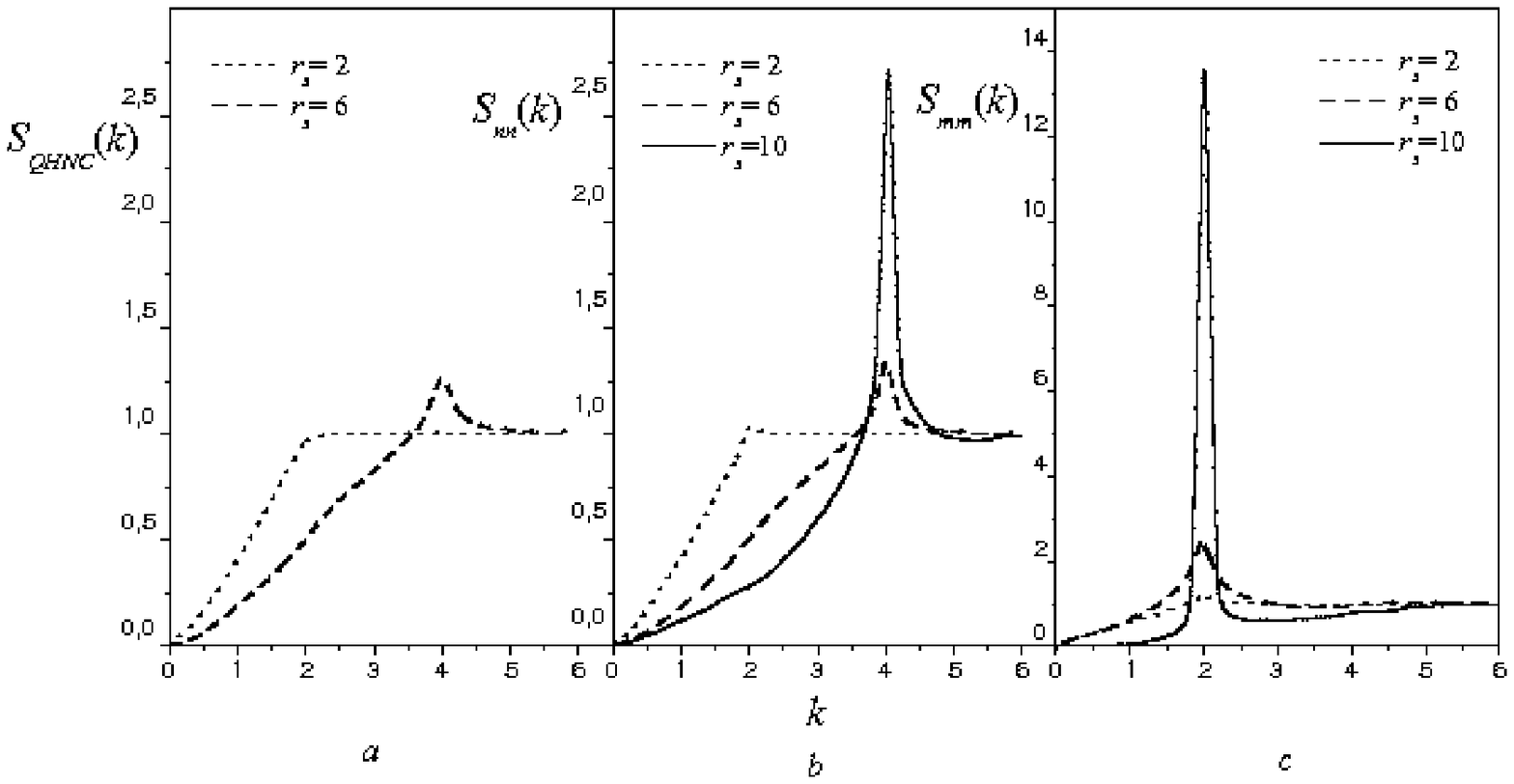}
 \caption{Structure factors for $b$ = 4 and diverse values of $r_{s}$. \textit{a}): QHNC; \textit{b}): VQMC numeric ;\textit{c}): VQMC magnetic. Momenta are reduced by the Fermi momentum.}
 \label{skb4}
\end{figure}

In figures \ref{skb01} and \ref{skb4} can be seen the structure factors, numeric (panel \textit{b}) and magnetic (panel \textit{c}), calculated from Monte Carlo simulations, for two values of the wire width  and diverse  densities. In the panel \textit{a} we show the structure factor (numeric) in the QHNC obtained by Fourier transforming $g(x)$. We mention that the lack of the corresponding curve for $r_{s}=10$ is because convergence in the solution of the QHNC equations was not reached in this case.

The most notable feature of curves is the peak located at $4\textit{k}_{F}$ in the numeric structure factor and at $ 2\textit{k}_{F}$ in the magnetic one.
This, as we will see below (eqs. \ref{w3} y \ref{w4}), says that, for small wire width the charge (or, in this case indistinctly, particle) correlations dominate over the spin correlations. At least for the densities we are considering, the effect seems to decrease when the width of the wire is increased, that is, when the system is more similar to a bidimensional one, and also when the density grows. This behavior is specially apparent in fig. \ref{skb4} \textit{a}, where can be clearly observed that for $r_{s}=2$ the peak moves at $\textit{k}=2\textit{k}_{F}$ in similar way as for electrons moving on a plane(\cite{senatore}). 

The peak at $\textit{k}=4\textit{k}_{F}$ observed mainly for small widths and densities, can be interpreted as indicating the system tendency towards a crystallization of the Wigner type. The wavelength corresponding to this value of $\textit{k}$ is $\lambda=\rho^{-1}=\frac{L}{N}$, say, the mean distance between particles.

Wigner crystallization in one dimensional quantum systems with long range interactions was firstly studied by Schultz (\cite{Schulz1})
by means of bosonization methods(\cite{Emery},\cite{Solyom}). The main conclusion was that long range forces, even if they are weak, cause a state characterized by a long range quasi-order more adequate to describe a one dimensional Wigner crystal than a liquid. 

The charge-charge correlations as estimated by Schultz is given by:
\begin{equation}
\left\langle \rho (x)\rho (0)\right\rangle =A_{1}\cos (2k_{F}x)\frac{e^{-c_{2}\sqrt{\ln x}}}{x}+A_{2}\cos (4k_{F}x)e^{-4c_{2}\sqrt{\ln x}}+...  
\label{w3}
\end{equation}
where $A_{1,2}$ are constants that depend on the interaction. The interesting point is the smooth variation
of the $4k_{F}$ term, smoother than any power law, showing an incipient charge density wave of wave number 
$4k_{F}$. As we have already mentioned, the period of the oscillations, $4k_{F}$, is the mean space between particles, that is, the expected value for a one dimensional Wigner crystal.

On the other hand, the spin-spin correlations are:

\begin{equation}
\left\langle S(x)S(0)\right\rangle \approx B_{1}\cos (2k_{F}x)\frac{e^{-c_{2}\sqrt{\ln x})}}{x}+..., 
\label{w4}
\end{equation}
where the lack of the $4k_{F}$ term should be noted. 

It is worth mentioning that the correlations of charge (eq. \ref{w3}) and spin (eq. \ref{w4}) are related with the numeric and magnetic structure factors, respectively, via Fourier transforms. 

\subsection{Quantum wires as semiconductors}
In the previous Subsection, quantum wires were taken as just a one dimensional electron gas, that is, as conductors seen from the Sommerfeld-Pauli point of view. Here we will consider them as semiconductors where the carriers are the electrons of the conduction band and the holes left by the electrons that jump from the valence band to the conduction one. Recall that in the effective mass approximation the semiconductors carriers effective mass depends on the band structure of the considered material and, in general, it will be different for electrons and holes.

We calculate the pair correlation functions using the QHNC equations for binary mixtures of Subsection \ref{QHNC_mixtures}.
In particular, we will consider the electron-hole pair correlations at contact and their relationship with the photoluninescence phenomenon experimentally observed in quantum wires. Thus the results that we will show can be taken as a true test for the QHNC theory against real experiments (not simulations)

Our system is constituted of electrons and holes with charges
$e_{-}=-e_{+}=e$, effective masses $m^{*}_{e}$ and $m^{*}_{h}$ and  densities $\rho_{e}=\rho_{h}=\rho$.

We assume that the particles interact through effective potentials of the form given by Eq.\ref{cc4}:

\begin{equation}
V_{eff}^{ij}(x)=e^{i}e^{j}\frac{\sqrt{\pi}}{2b} \exp \left(\frac{x^{2}}{4b^{2}} \right) \text{erfc} \left( \frac{\vert x \vert }{2b} \right)
\hspace*{2cm}i,j =e,h
\end{equation}
where, as before, $b$ denotes the wire width.

\subsubsection{Contact pair correlation and photoluminescence}

Photoluminescence phenomena are of interest in relation with photodetectors and laser devices. Some features of photoluminescence in semiconductors can be explained in terms of the radiative recombination of electrons and holes: an electron of the conduction band decays into the valence band in indirect form and annihilate with a hole with the correspondent emission of radiation.

The electron-hole recombination rate in a system as the one we are considering here is considerably higher than in an hypothetical ideal system in which the carriers do not interact among them. This increase in the recombination rate is usually assigned to an enhancement factor $\textit{g}_{eh}(0)$, say the enhancement factor is rightly the electron-hole correlation at contact.

Since the correlation function at contact $g_{eh}(x=0)$ is a measure of the probability that an electron meets a hole and the photoluminescence intensity is inversely proportional to the recombination time $\tau $, then the photoluminescence intensity can be studied in terms of the contact correlation functions by using the expression \cite{Rice1}.

\begin{equation}  \label{63}
\frac 1\tau =\frac 1{\tau _0}g_{eh}(0)
\end{equation}
where $\tau _0$ is the  radiative recombination rate for the electron-hole pair when interactions are turned off.

\begin{figure}[here]
 \centering
 \includegraphics[width=1.0\textwidth]{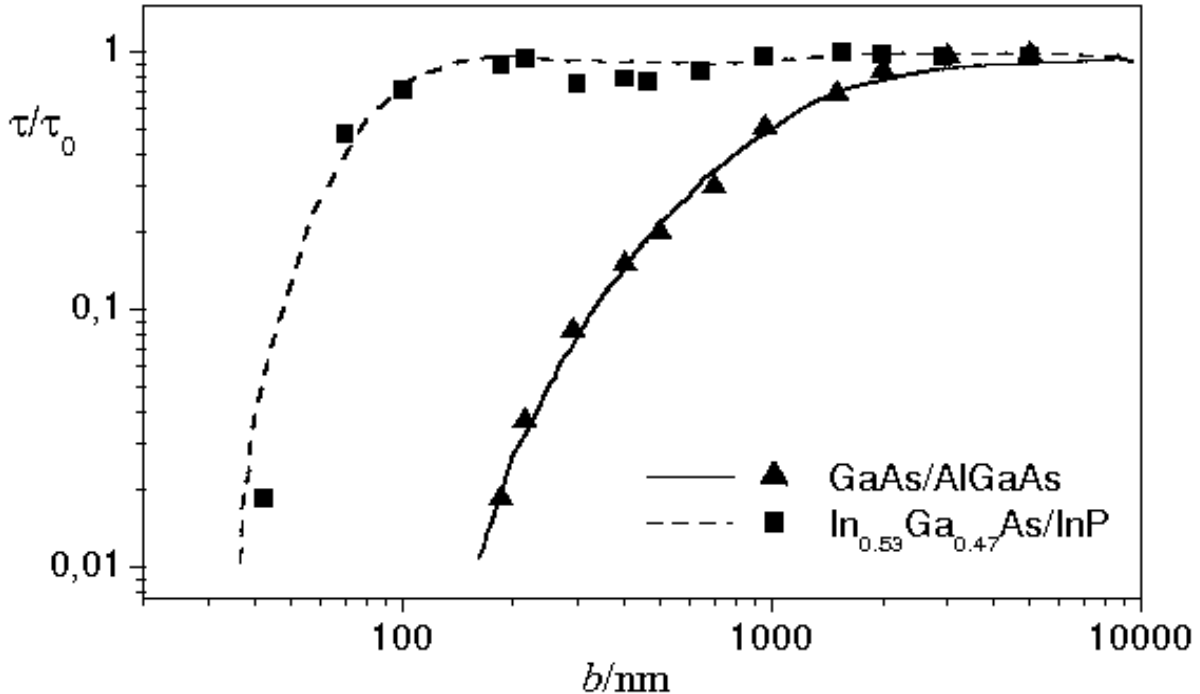}
 \caption{Comparison of the curves $\tau$ vs. $b$ as calculated with QHNC approximation (lines) and those obtained from photoluminescence experiments in quantum wires and reported in ref.\cite{Pilkuhn1} (symbols).} 
 \label{fotoluminiscencia}
\end{figure}
In fig. \ref{fotoluminiscencia} show the QHNC results for the mean life time $\tau$ of the electron-hole pairs as a function of the width
$b$ for two quatum wires: $GaAs/AlGaAs$ y $In_{0.53}Ga_{0.47}As/InP$. Also the corresponding experimetal points obtained by a group of Sttutgart University are shown\cite{Pilkuhn1}.

The effective masses we use are $m_{e}^{\ast }$ = 0.067 $m_{e}$; $m_{h}^{\ast }$ = 0.45 $m_{e}$ for $GaAs/AlGaAs$\cite{Sze1} and $m_{e}^{\ast
}$ = 0.041 $m_{e}$; $m_{h}^{\ast }$ = 0.5 $m_{e}$ for $In_{0.53}Ga_{0.47}As/InP$ \cite{Razeghui1}. 
The one dimensional electronic density was obtained by fitting an experimental point. the resulting values: $n_{e}$ = 1.47$\times 10^{6}[cm^{-1}] f$ 
for $GaAs/AlGaAs$ and $n_{e}$ = 1.85 $\times 10^{6}$ $[cm^{-1}]$ for $In_{0.53}Ga_{0.47}As/InP$, are reasonable for these systems.

We can take the good agreement between theoretical and experimental results as a validation of our QHNC version to study experimental realizations of one dimensional fermionic systems. It should be mentioned that perturbation calculations within ladder approximation previously performed in our group yield similar results\cite{Vericat1},\cite{Vericat2} 

\subsection{Coupled quantum wires}

\label{cca} 
Up to now we have considered just an isolated quantum wire. However we have already pointed out that the manufacture process yields a series of nearly parallel wires which are separated by nanometric distances. As a first approximation to this scenario we study a couple of parallel wires. To this we use again the extension of the QHNC approximation we have developed for binary mixtures (Subsection \ref{QHNC_mixtures}). 
To fix ideas we denote with $e$ the carriers of one of the wires and with $h$ those of the other one. Here we will use Schultz potential (eq. \ref{schultz}). For fermions of the same wire the effective pair interaction is:

\begin{equation} 
 V_{eff}^{ii}(x)=\dfrac{(e^{i})^{2}}{\sqrt{x^{2}+b^{2}}},
\hspace*{2cm} i=e,h
\end{equation} 
whereas between carriers of different wires the interaction will be:
\begin{equation} 
 V_{eff}^{ij}(x)=\dfrac{e^{i}e^{j}}{\sqrt{x^{2}+b^{2}+d^{2}}},
\hspace*{2cm} i\neq j=e,h
\label{schultzab}
\end{equation} 

\begin{figure}
 \centering
 \includegraphics[width=1.0\textwidth]{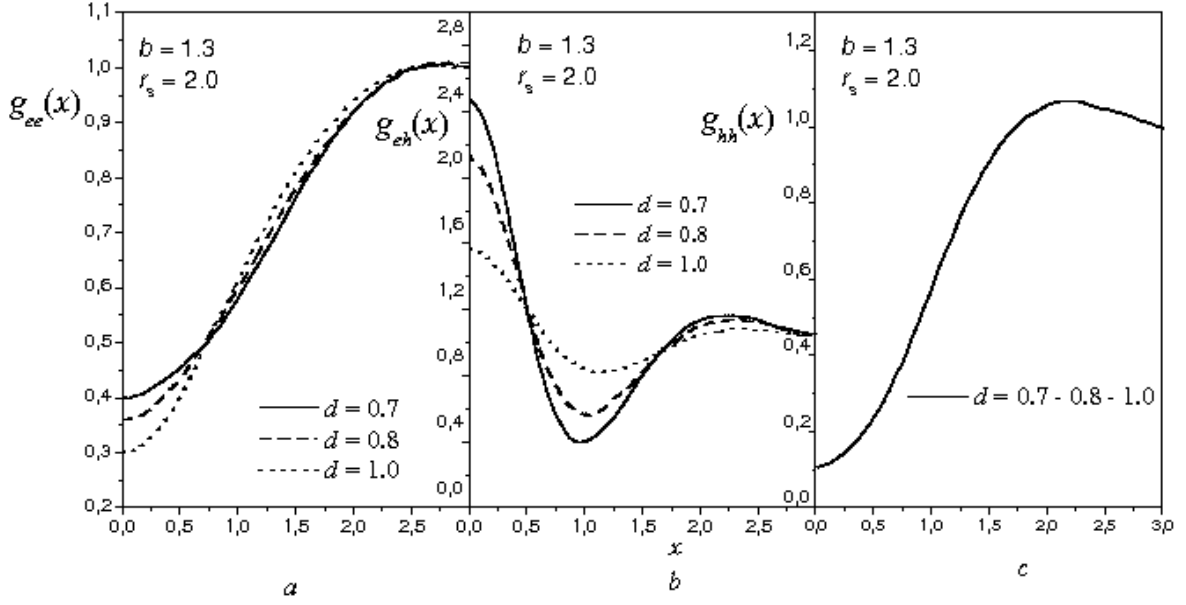}
 \caption{Pair correlation functions in the QHNC approximation for coupled wires of reduced width $b$ = 1.3 at a fixed density ($r_{s}=2$) and diverse distance between the wires. \textit{a}) electron-electron; \textit{b}) electron-hole; \textit{c}) hole-hole.} 
 \label{gacoprs2}
\end{figure}
\begin{figure}
 \centering
 \includegraphics[width=1.0\textwidth]{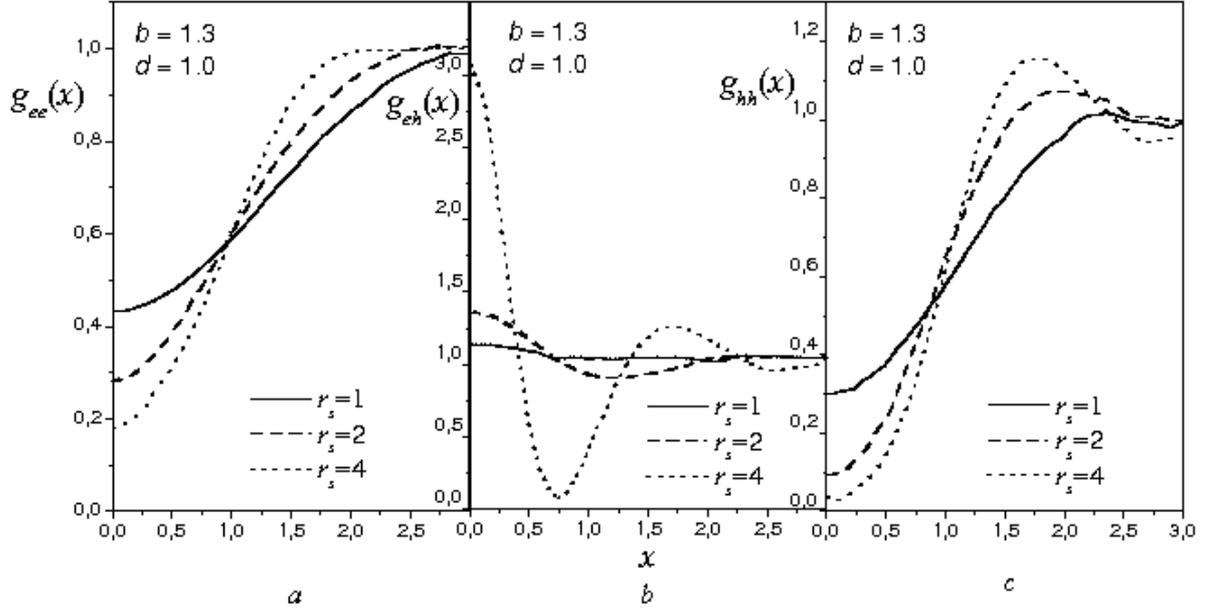}
 \caption{Pair correlation functions in the QHNC approximation for coupled wires of reduced width $b$ = 1.3 at a fixed distance between the wires ($d=1$) and diverse densities. \textit{a}) electron-electron; \textit{b}) electron-hole; \textit{c}) hole-hole.} 
 \label{gacopd1}
\end{figure}

Fig. \ref{gacoprs2} shows the pair correlation functions inter and intra wires for the same density and distinct distances between wires, whereas in fig.\ref{gacopd1} can be observed the same functions but for different densities when the distance remains constant. In these figures the distance are in units of the inverse of the Fermi momentum. The ratio of the effective masses to the electron mass are
$\frac{m^{*}_{h}}{m^{*}_{e}}=7$ and wire diameter in units of the Bohr radius $a_{0}=\frac{\hslash^{2}\epsilon}{m^{*}_{e}e^{2}}$ is $b/a_{0}$ =1.3. We are taking a dielectric constant $\epsilon=13$.
In relation with these two figures it must be understood that for the functions $g_{eh}(x)$ (panel \textit{b}) the origin corresponds to a distance $d$ (minimum distance between wires).

From fig. \ref{gacoprs2} \textit{a} y \textit{b} we conclude that, at least for the values we are considering here, the variation in the distance between wires hardly affects the correlations inside a given wire. As far as $g_{eh}(x)$ (panel \textit{b}) its variation with $d$ is more noticeable. This is reasonable because of the dependence of the effective pair potential with the distance between wires (see eq. \ref{schultzab}).

Also from fig. \ref{gacopd1} we see that, for a fixed distance between wires, the correlations inside a given wire (panels \textit{a} and \textit{c}) depend on the density as it would be expected; the greater the density the smaller the structure as for a degenerate gas. On the other hand the curves for $g_{eh}(x)$ show as remarkable feature a noticeable increasing at contact ($x=d$) when density decreases. This fact can be interpreted as the indication of an exciton formation.

Similar QHNC calculations but for coupled electron and hole quantum wells have been reported by Alatalo \textit{et. al.} \cite{Alatalo1}, \cite{Alatalo2}, \cite{Alatalo3}. More specifically, Yurtsever and Tanatar \cite{Yurtsever1} studied coupled quantum wire systems using a perturbation approach based in the ladder approximation of Yasuhara \cite{Yasuhara1}.

\section{Conclusions}

In this work we have tackled the problem of describing one dimensional systems
of many fermions. To this end we have developed a realization of
Fantoni-Rosati formalism, which yields a version of the hypernetted chain
approximation that shows, with respect to other versions available in the
literature, some remarkable aspects, particularly the way in which the
distinct classes of graphs are ordered and summed and also the form considered
for the energy variational equation. We apply our QHNC equations to study
quantum wires modeled, according to Sommerfeld-Pauli picture, as a one
dimensional electron gas or as a mixture of electrons and holes in 1D. The
confinement effects are taken into account through one dimensional pair
potentials that include as a parameter the wire width. Our QHNC results were
compared with variational Monte Carlo simulations and, in the case of
electron-hole mixtures, also with photoluminescence experiments in quantum
wires.
\begin{acknowledgments}
Support of this work by Consejo Nacional de Investigaciones Cient\'{\i}ficas
y T\'{e}cnicas (PIP 112-200801-01192 ), Universidad Nacional de La Plata\
(Grant 11/I108), Universidad Nacional de Rosario (Grant 19/J089) and 
Agencia Nacional de Promoci\'{o}n Cient\'{\i}fica y
Tecnol\'{o}gica of Argentina (PICT 2007-00908) is greatly appreciated. 
C.M.C and F.V. are members of CONICET.\ 
\end{acknowledgments}

\end{document}